\documentclass[useAMS,usenatbib]{mn2e}
\usepackage{graphicx}
\usepackage{latexsym}
\begin{document}

\title[Dusty Mg\,{\normalsize\it II} absorbers]{Dusty Mg\,{\Large\bf II} absorbers: population statistics, extinction curves and gamma-ray burst sightlines}

\author[J. M. Budzynski and P. C. Hewett]{J. M. Budzynski\thanks{E-mail: jbudzyn@ast.cam.ac.uk} and P. C. Hewett\\
Institute of Astronomy, University of Cambridge, Madingley Road, Cambridge, CB3 0HA, UK\\}

\pagerange{\pageref{firstpage}--\pageref{lastpage}} \pubyear{2010}

\newcommand{\dd}{\textrm{d}}
\newcommand{\rin}{R_\textrm{\tiny{in}}}
\newcommand{\OO}{\mathcal{O}}
\maketitle

\label{firstpage}

\def\mnras{MNRAS}
\def\memras{MEMRAS}
\def\apj{ApJ}
\def\aap{A\&A}
\def\apjl{ApJL}
\def\apjs{ApJS}
\def\araa{ARA\&A}
\def\nat{Nature}
\def\aj{AJ}
\def\pasp{PASP}
\def\apss{Ap\&SS}

\def \kms {\rm \,km\,s$^{-1}$}

\newcommand{\mg}{Mg\,{\small II}}  
\newcommand{\sithree}{Si\,{\small III}}
\newcommand{\hone}{H\,{\small I}}
\newcommand{\nfive}{N\,{\small V}}
\newcommand{\sitwo}{Si\,{\small II}}
\newcommand{\oone}{Si\,{\small III}}
\newcommand{\ctwo}{C\,{\small II}}
\newcommand{\sifour}{Si\,{\small IV}}
\newcommand{\fetwo}{Fe\,{\small II}}
\newcommand{\altwo}{Al\,{\small II}}
\newcommand{\althree}{Al\,{\small III}}
\newcommand{\mgone}{Mg\,{\small I}}
\newcommand{\mgcap}{Mg\,{\sc ii}~}
\newcommand{\caii}{Ca\,{\small II}}
\newcommand{\ang}{\,\AA~}

\newcommand{\basenum}{8\,329} 
\newcommand{\uncertnum}{4\,550} 
\newcommand{\compnum}{211}
\newcommand{\mwnum}{17}
\newcommand{\smcspec}{137}  
\newcommand{\lmcspec}{74}
\newcommand{\mwspec}{0}
\newcommand{\numhighebv}{25}
\newcommand{\numextreme}{2} 
\newcommand{\highfrac}{13} 
\newcommand{\totfrac}{12} 
\newcommand{\plnorm}{\left(8.0 \pm 3.0\right)\times 10^{-4}}
\newcommand{\plpower}{\left(3.48 \pm 0.3\right)}
\newcommand{\grbdisc}{1.24 \pm 0.05}
\newcommand{\mgfrac}{36}
\newcommand{\mgmissperc}{24}
\newcommand{\mgmisspercerr}{4}
\newcommand{\mgmissperchigh}{34}
\newcommand{\mgmissperchigherr}{2}


\begin{abstract}
We present a new determination of the dust content and
near-ultraviolet/optical extinction curves associated with a sample of
$\simeq$8300 strong, $W_{0}^{\lambda2796}>1$\,\AA, \mg \ absorbers,
redshifts 0.4$<$$z$$<$2.2, identified in Sloan Digital Sky survey
(SDSS) spectra of quasars in the DR6 release. Taking into account the
selection effects that result from dust extinction, including the
reduction in the signal-to-noise ratio of an absorber appearing in a
reddened quasar spectrum, we find a stronger dependence of $E(B-V)$ on
absorber $W_{0}^{\lambda2796}$ than in other published work. The
dependence of the median reddening on $W_{0}^{\lambda2796}$ can be
reproduced by the power-law model
$E(B-V)=\left( \plnorm \right) \times \left(W_0 \right) ^{ \plpower }$ for
1.0$\le$$W_0$$\le$5.0.  Observed \mg \ samples, derived from
flux-limited quasar surveys, are shown to suffer from significant
incompleteness at the level of \mgmissperc \ $\pm$ \mgmisspercerr \
per cent for absorbers with $W_0>1$\ang and \mgmissperchigh \ $\pm$
\mgmissperchigherr \ per cent for absorbers with $W_0>2$\,\AA. Direct
determination of the shape of the near-ultraviolet extinction curves,
using high signal-to-noise ratio composites, for absorbers as a
function of $E(B-V)$ show evidence for systematic changes in the form
of the extinction curves. At low $E(B-V)$ ($\la$0.05), the extinction
curve is featureless and well represented by a Small Magellanic Cloud
(SMC)-like extinction curve. For intermediate $E(B-V)$s ($\la$0.2),
approximately a third of \mg \ absorbers show evidence for a 2175\ang
feature and an extinction curve similar to that of the Large
Magellanic Cloud (LMC). For the small number of high $E(B-V)$
($\ga$0.3) absorbers, the majority of which exhibit strong \caii
$\lambda\lambda$3935,3970 absorption, there is evidence for the
presence of a 2175\ang feature as strong as that found in the Milky
Way (MW). Near-infrared photometry for six of the systems indicates
that the rest-frame optical portion of the extinction curve for these
high-$E(B-V)$, and likely very high column density, systems is
significantly greyer than the SMC, LMC or MW extinction curves.
Application of the new results on the dust content of strong \mg \
absorbers shows that dusty absorbers can account for a significant
proportion, up to a factor of two, of the observed overdensity of
absorbers seen towards Gamma-Ray Burst (GRB) sightlines, compared to
sightlines towards quasars in flux-limited samples.

\end{abstract}


\section{Introduction}\label{intro}
Since the pioneering work of \citet{1986A&A...155L...8B},
\citet{1990ApJ...357..321L, 1992ApJ...391...48L} and
\citet{1994ApJ...437L..75S} the presence of \mg
\ $\lambda\lambda$2796,2803 absorbers in `haloes' extending to
distances of $\sim$50--100\,$h^{-1}$ kpc about luminous galaxies has
been well established. The more recent recognition of the importance
of outflow, infall and feedback processes in general for our knowledge
of galaxy evolution has reinvigorated attempts to understand the
processes responsible for the existence of extended gaseous haloes
associated with luminous galaxies.

Strong \mg \ absorbers, \mg \ $\lambda$2796 rest-equivalent width (EW)
$\ga$0.5\,\AA, seen in the spectra of background quasars reveal the
presence of relatively cool, $T$$\sim$10$^4$, ionised gas with neutral
hydrogen column densities of
$\simeq$10$^{18}$--10$^{22}$\,cm$^{-2}$. The availability of very
large samples of intermediate resolution, moderate signal-to-noise
ratio (S/N) quasar spectra from the Sloan Digital Sky Survey (SDSS)
\citep{2000AJ....120.1579Y}, has allowed the compilation of extended
samples of \mg \ absorbers \citep[e.g.][]{2005ApJ...628..637N,
  2006ApJ...639..766P,2011AJ....141..137Q} with consequent
improvements in our knowledge of the statistical properties of the
absorbers as a function of key physical parameters.
     
The homogeneity of the SDSS spectroscopy coupled with the quantitative
nature of the quasar target selection \citep{2002AJ....123.2945R} has
also enabled progress to be made in quantifying the dust content of
absorber populations at cosmological
distances. \citet{2005MNRAS.361L..30W} first demonstrated that the
\caii \ $\lambda\lambda$3935,3970 absorber population, a subset of
very strong \mg \ absorbers, showed evidence for the presence of dust
with $E(B-V)$ values as high as 0.2 magnitudes. The strong \mg
\ absorber population as a whole shows much lower dust content, with
mean $E(B-V)$$\simeq$0.02 \citep{2006MNRAS.367..211W,
  2006MNRAS.367..945Y}.  More recently, \citet{2008MNRAS.385.1053M}
quantified the dependence of the dust content of \mg \ absorbers as a
function of both absorber EW and redshift.

Interest in the properties of dust associated with \mg \ absorbers has
grown with the explicit linking of the absorber population to the star
formation history of the galaxy population \citep{2009arXiv0912.3263M}
and the association between very strong absorbers and outflows
\citep{2007NewAR..51..131B, 2010arXiv1003.0693N}.  Dust has also been
suggested as a possible explanation for the puzzling discrepancy
between the increased incidence of \mg \ absorbers detected in
Gamma-Ray Burst (GRB) compared to quasar sightlines
\citep{2006ApJ...648L..93P, 2007ApJ...659..218P}.

In the local Universe the principal observable difference between dust
extinction curves is the presence of the 2175\,\AA \ feature, the
strength of which is inversely correlated with the increasing rise of
dust extinction into the ultra-violet. The 2175\,\AA \ feature is
strong within the Milky Way (MW) but only weakly detected within the
Large Magellanic Cloud (LMC) and is practically absent within the
Small Magellanic Cloud (SMC). The origin of the feature is generally
believed to be aromatic carbonaceous materials, i.e. a mixture of
Polycyclic Aromatic Hydrocarbon (PAH) molecules, which are abundant in
the MW \citep{2003ARA&A..41..241D, 2005pgqa.conf..331W}. A few reports
of the detection of a strong 2175\,\AA \ feature in individual
absorbers have been made, e.g. \citet{2008MNRAS.391L..69S} and
\citet{2010ApJ...724.1325J} analysing two individual \mg \ systems
each, \citet{2010ApJ...708..742Z} analysing a single extremely
dusty object, and very recently \citet{2011ApJ...732..110J}, who report 39 detections.
On the other hand, statistical studies of both \caii
\ \citep{2005MNRAS.361L..30W} and \mg \ absorbers
\citep{2005pgqa.conf..331W,2006MNRAS.367..945Y} favour SMC- or
LMC-like extinction curves. Such studies to date have been
characterised by modest sample sizes and the $E(B-V)$ values of the
absorbers are typically small, which means the limits on the presence
of any sub-population of absorbers possessing MW-like extinction
curves are not strong.

In this paper we employ a new, large, sample of \mg \ absorbers,
derived using quasars contained in the SDSS DR6-release, to
investigate the form of the rest-frame extinction curve associated
with the absorber population with unprecedented accuracy.  Adopting
the approach of Wild et al. (2006), a full analysis of the presence of
dust in the absorbers allows a quantitative determination of the
selection biases that effect the observed \mg \ absorber sample. With
such a quantification in hand we then review the implications of the
selection biases on i) the redshift evolution of the dust content of
\mg \ absorbers, ii) the redshift distribution of very strong \mg \ 
absorbers, and iii) the predicted number of \mg \ absorbers along
GRB-sightlines.

The outline of this paper is as follows. In Section \ref{mgsample} we
describe the quasar and \mg \ absorber catalogues which are used in
Section \ref{dustquant} to estimate statistically the amount of dust
in individual \mg \ systems. We also assess the effect of a dust bias on
the completeness of the sample in Section \ref{sec:dustobsbias}. In
Section \ref{analyres} we show results of an investigation to identify 
the 2175\ang bump in the strongest \mg \ systems and we look at the
correlations between absorber EW and redshift. We also discuss the
implications of the results for \mg \ absorber statistics and the
discrepancy in the observed frequency of absorbers along GRB- and
quasar-sightlines in Section \ref{analyres}, before summarising our
conclusions in Section \ref{disc}.
Throughout this work we assume a cosmology of $\Omega _{\mathrm{M}}=0.3$, $\Omega _{\mathrm{\Lambda}}=0.7$, and $h=0.7$.

\section{Quasar and M\lowercase{g}\,{\sevensize\bf II} 
absorber samples}\label{mgsample}

\begin{figure*}
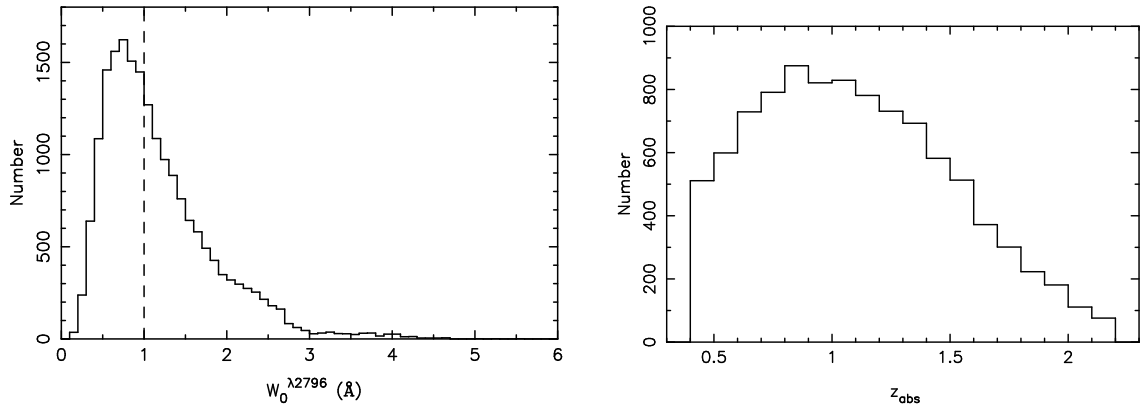

\includegraphics[width=0.3\textwidth,angle=270]{jb_fig1.ps}
\hspace{5mm}
\includegraphics[width=0.3\textwidth,angle=270]{jb_fig2.ps}
\caption{Left-hand panel: the equivalent width (EW of the \mgcap
  $\lambda2796$ line) distribution of absorbers in a sample of 19\,315
  systems from SDSS. The vertical dashed line at 1\ang corresponds to
  lower EW-limit of absorbers retained in our sample (9719
  systems). The sample is highly complete down to EW$=1$\,\AA.
  Right-hand panel: the redshift distribution for systems with
  EW$\ge1$\,\AA.}
\label{cap:newz}
\end{figure*}

The base quasar sample consists of 91\,665 objects, including 77\,392
quasars in the \citet{2007AJ....134..102S} DR5 catalogue that are
retained in the later DR7 quasar catalogue of
\citet{2010AJ....139.2360S}.  A further 13\,081 objects are quasars,
present in the additional DR6 spectroscopic plates, identified by one
of us (PCH) using a similar prescription to that employed by
\citet{2007AJ....134..102S}, all of which are present in the
\citet{2010AJ....139.2360S} catalogue.  An additional 1192 objects,
which do not satisfy one, or both, of the emission line velocity width
or absolute magnitude criterion imposed by
\citet{2007AJ....134..102S}, are also included.  None of the results
in the paper depend on the exact definition of the `quasar'-sample
used.

The spectra are all processed through the sky-residual subtraction
scheme of \citet{2005MNRAS.358.1083W}, resulting in significantly
improved S/N at wavelengths $\lambda$$>$7200\ang and consequent
increased detection efficiency for \mg \ absorbers at $z$$>$1.6.

The quasar sample is restricted to the redshift interval 0.5$\le$$z
$$\le$3.5 and to objects with Galactic extinction corrected SDSS i-band
magnitudes $m_i\le19.1$, leaving 52\,112 quasars.  Broad Absorption Line
(BAL) quasars identified by \citet{2009ApJ...692..758G} or from our
own BAL catalogue \citep{2011MNRAS.410..860A} are also
excluded, leaving 48\,587 quasars.

The presence of extended wavelength intervals without data in the SDSS
quasar spectra is not ideal for the extinction estimation
(Section~\ref{dustquant}). The sample is therefore restricted to
essentially `complete' spectra by the requirement that the number of
valid pixels, {\tt NGOOD}$>$3500, eliminating a further 490 quasars.

The final sample of 48\,097 quasars is searched for \mg \
$\lambda\lambda2796.35,2803.53$ absorption doublets in the redshift
interval 0.4$\le$$z_{\mathrm{abs}}$$\le$2.2.  A `continuum' is defined
for each quasar via the application of a simple 61--pixel median
filter.  The `difference' spectrum, to be searched for absorption
features, is then obtained by subtracting the continuum from the
original quasar spectrum.  The absorption line search uses a
matched-filter technique \citep[e.g.][]{1985MNRAS.213..971H} with two
template Gaussian doublets of the appropriate wavelength separation
and full width at half maximum (FWHM) = 160\kms (the resolution of the
SDSS spectra), 200\kms and 240\kms.  Three values of the doublet ratio
are incorporated into the search: 2:1 (corresponding to unsaturated
lines on the linear part of the curve of growth), 1:1 (for saturated
lines on the flat part of the curve of growth) and an intermediate
case, 4:3.  A further template with doublet ratio 1:1 and
flat-bottomed absorber profiles, 380\kms in extent, is also used to
optimise the detection of very high EW doublets.  At each pixel the
template giving the minimum $\chi2$ value is determined and candidate
absorbers selected by applying a threshold value of S/N$\ge$7$\sigma$.

A total of 19\,315 \mg \ systems are indentified but we confine our
analysis to the sub-sample of 9\,719 systems with EW of the
2796\ang \ line ($W_0^{\lambda2796}$)$\ge$1\AA.  The left-hand
panel of Fig.~\ref{cap:newz} shows the distribution of
$W_0^{\lambda2796}$ with the $W_0^{\lambda2796}$=1\AA-limit indicated.
The right-hand panel of Fig.~\ref{cap:newz} shows the redshift
distribution for the $\ge$1\,\AA \ sample.  Providing that any dust
associated with the \mg \ absorber population results in an extinction
curve whose shape is not strongly dependent on $W_0^{\lambda2796}$,
the investigation presented here is not sensitive to the completeness
of the absorber catalogue.  However, the statistical properties of the
absorber catalogue are in excellent agreement with previous work
\citep{2005ApJ...628..637N}.  The catalogue is highly complete down to
$W_0^{\lambda2796}$=1\,\AA \ and, from a combination of visual
inspection and investigation of the properties of other absorption
species in `stacked' \mg \ systems, contamination by false detections
is found to be $<$5 per cent.

There are 7\,318 quasars with a single intervening \mg \ absorber,
1\,019 quasars with two absorbers and 116 quasars with more than two
absorbers. Due to the increased uncertainty in the determination of
the dust content of individual absorption systems in quasars
containing multiple absorbers (i.e. $> 2$ systems), we chose to retain
only quasars with single or double intervening \mg \ absorbers. In the
case of a double absorber we consider only the stronger system.  The
vast majority of double absorbers involve one or more low EW systems,
however, we remove from the sample the $8/1019$ spectra that possess
two absorbers with an EW$>2.5$\ang \ to leave a total of 8\,329
systems. The results and conclusions derived in the paper are not
sensitive to whether we use only single systems or single plus double
systems.

\begin{figure*}
\includegraphics[width=0.4\textwidth]{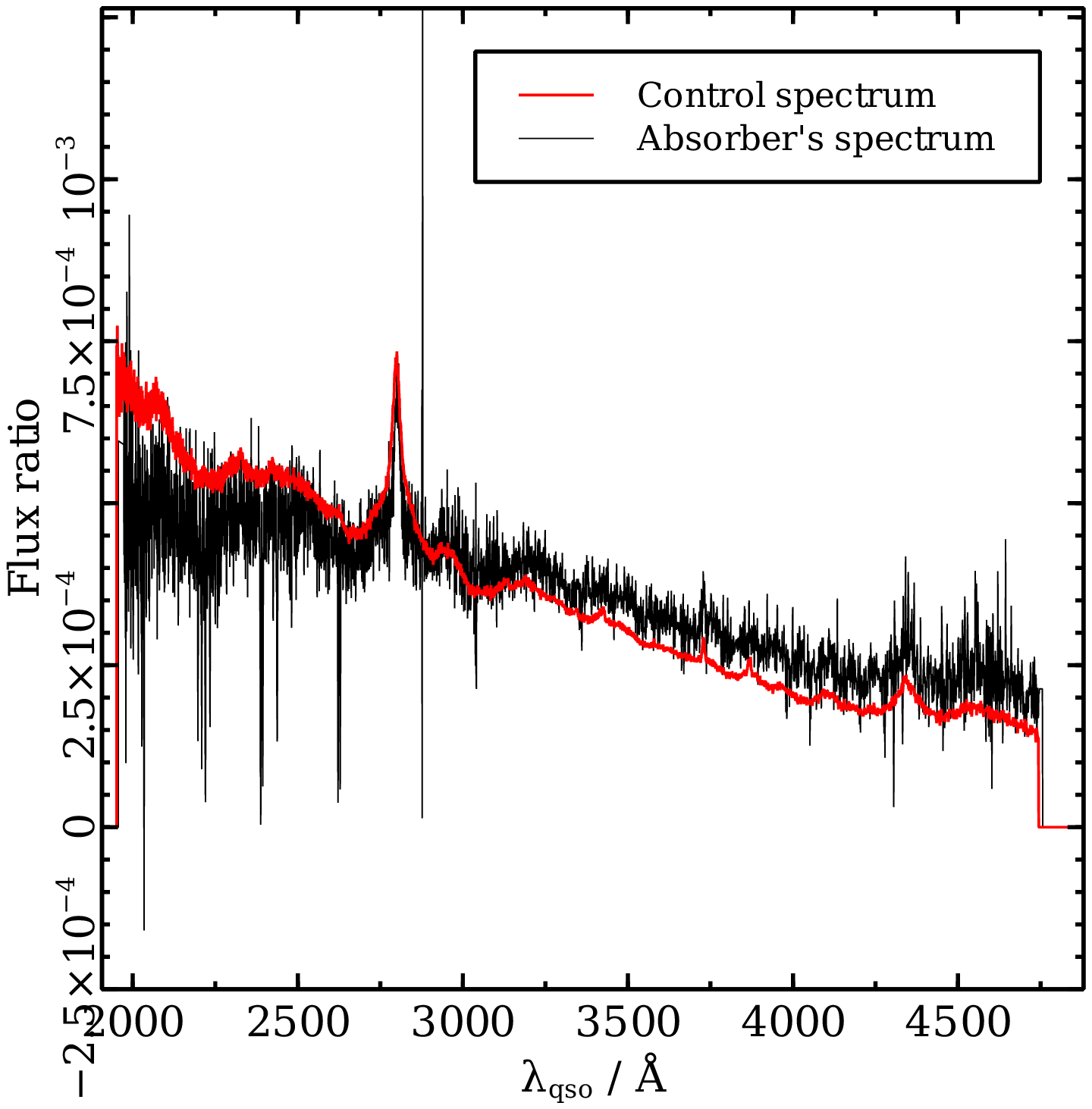}
\hspace{5mm}
\includegraphics[width=0.4\textwidth]{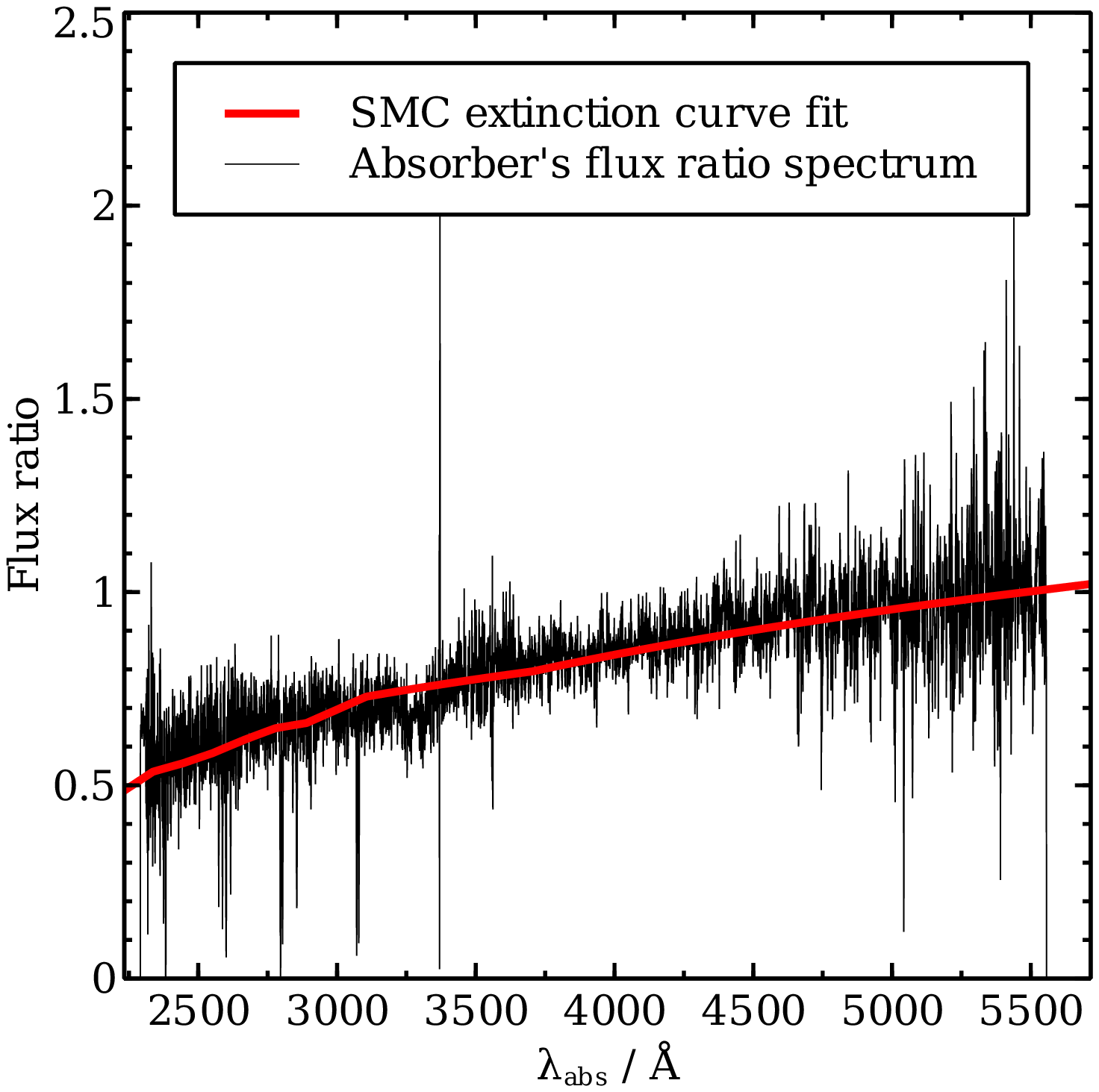}
\caption{Left-hand panel: flux ratio as a function of wavelength for
  quasar SDSS J093508.37+271648.6, $z_{\mathrm{qso}}=0.94$, with an
  \mg \ absorber at $z_{\mathrm{abs}}=0.66$. A `control' quasar
  spectrum, representing the spectrum of the background quasar, is
  overplotted in red. Right-hand panel: flux ratio spectrum for the
  $z_{\mathrm{abs}}=0.66$ \mg \ absorber in the rest-frame of the
  absorber, obtained by dividing the quasar spectrum by the control
  spectrum in the left-hand panel. The red line shows the best fitting
  SMC extinction curve, which corresponds to an $E(B-V)=0.15$.}
\label{cap:medianext}
\end{figure*}

\section{Measuring the dust content of the absorbers}\label{dustquant}
The presence of dust in an \mg \ absorber affects the observed
spectrum of the background quasar. The degree of reddening of a
spectral energy distribution (SED) is described by an associated
extinction curve, and the dust content is parameterised by a
corresponding $E(B-V)$\footnote{$E(B-V)$ values refer to the absorber
  rest-frame unless specifically indicated otherwise.}
(e.g. \citet{1994ApJ...429..172K}). The large and homogenous SDSS DR6
spectroscopic catalogue provides an opportunity to obtain estimates of
the dust content for many thousands of absorbers. However, one must be
sure that intrinsic quasar-to-quasar SED variations do not produce
significant systematic biases in, or add significant scatter to, the
$E(B-V)$ determinations of the absorbers.

The population of \mg \ absorbers is reported to have a mean dust
content of only $E(B-V)\simeq0.02\,-\,0.03\,\mathrm{mag}$
\citep{2006MNRAS.367..211W,2006MNRAS.367..945Y}, which implies only
small alterations to the shape of the SEDs of background
quasars. Therefore, care must be taken to construct a quasar `control'
spectrum, which represents accurately the spectrum of an unabsorbed
quasar. Such a spectrum then enables us to isolate the effect of dust
in an absorber on the background quasar spectrum.

\subsection{The control quasar spectrum}\label{sub:controlSpectrum}
We have adopted an approach similar to \citet{2005MNRAS.361L..30W} and
\citet{2006MNRAS.367..945Y}, which involves using the two-dimensional
redshift, $z$, versus magnitude, $m_i$, plane to identify a sample of
quasars from which to construct a control spectrum. The use of
`neighbour' quasars with similar redshifts in the control sample
ensures that the wavelength coverage of control quasars is nearly
identical to that of the target quasar, while also ensuring that the
control sample quasars do not have systematically different (redshift
dependent) SEDs. At fixed redshift the quasar apparent magnitude
distribution is equivalent to the absolute magnitude distribution,
$M_i$. The large density of objects in the $m_i$ versus $z$ plane allows the
control sample to be defined using magnitudes close to that of the
target quasar, minimising the effect of any systematic
magnitude-dependent SED changes.

The control spectrum was generated for each target quasar, by taking
the median value (at each wavelength) of the 50 immediately brighter
and 50 immediately fainter quasar spectra within a fixed redshift interval
of $\Delta z=0.1$ centred on the redshift of the target
quasar\footnote{The 100 quasar spectra were shifted to the quasar rest-frame,
and normalised over the wavelength interval common to all spectra. The
median value at each rest-frame wavelength was then calculated.}. The
median was chosen instead of the mean, as it was less sensitive to
flux outliers arising from intrinsic variation among quasar SEDs. A
control spectrum is created for each of the 9\,719 quasars with
associated \mg \ absorbers, but only quasars without a detected
\mg \ absorber are allowed to contribute to the control spectra.

Using a fixed number of quasars to define the control spectrum, rather
than a fixed magnitude interval, means that the control sample is not
biased due to the systematic mismatch between the luminosity of the
target quasar and median luminosity of the control quasars.

\subsection{Estimation of the $\bmath{E(B-V)}$ values for the 
absorbers}\label{sub:Determining-E(B-V)}
The flux ratio for an absorber is calculated by dividing the target
quasar spectrum by the control spectrum in the quasar rest-frame,
removing, statistically, the signature of the quasar SED. The
resulting flux ratio spectrum is then moved to the absorber
rest-frame (Fig.~\ref{cap:medianext}).

We have used a parameterisation of the MW, LMC and SMC 
extinction curves, due to \citet{1992ApJ...395..130P}, and fit a curve of 
the form:

\begin{equation} 
F\left(\lambda\right)=F_{0}\,10^{-\frac{A\left(\lambda\right)}{2.5}},\label{eq:fluxebvfit}
\end{equation}

\noindent to the flux ratio spectrum, where $A\left(\lambda \right)$ is
the extinction at a particular wavelength given by

\begin{equation} 
A\left ( \lambda \right )=E\left ( B-V \right )\left[ \frac{E_{ \lambda - V }}{ E_{B - V}}  + R_{V}\right],\label{eq:alamdef}
\end{equation}

\noindent where $E_{ \lambda - V }/ E_{B - V}$ is given by
  \citeauthor{1992ApJ...395..130P}, and $R_{V}$ is the ratio of total
  to selective extinction. We have adopted $R_{v}=3.0$, which is
  representative of the value in the MW and SMC\footnote{The commonly
    adopted values of $R_{V}$ in the SMC and MW are 3.1 and 3.2
    respectively \citep{2007ApJ...663..320F}. The shape of the
    extinction curve does \emph{not} depend on the value of $R_{v}$.}.
  $F_{0}$ and the reddening parameter $E(B-V)$ are left as free
  parameters in the fit, which is carried out using the non-linear
  `Levenburg-Marquardt' algorithm \citep{marquardt:431}. The
  wavelength intervals corresponding to the strong telluric sky
  emission lines at $\lambda\lambda5578.5,6301.7$, and to the
  prominent absorption lines\footnote{The wavelengths of masked
    absorption lines are: \fetwo(1527, 1608, 2250, 2261, 2344, 2374,
    2383, 2587, 2601\,\AA), \altwo(1671\,\AA), \althree(1855,
    1863\,\AA), \mgone(2853\,\AA), \mg(2796, 2803\,\AA) and
    \caii(3935,3970\,\AA). The mask consists of 10\,\AA \ wide intervals
    centred on the absorption lines.} of \fetwo, \altwo, \althree,
  \mgone, \caii, which may be present in the absorber spectrum, are
  excluded from the fit.

\begin{figure}
\includegraphics[width=\columnwidth]{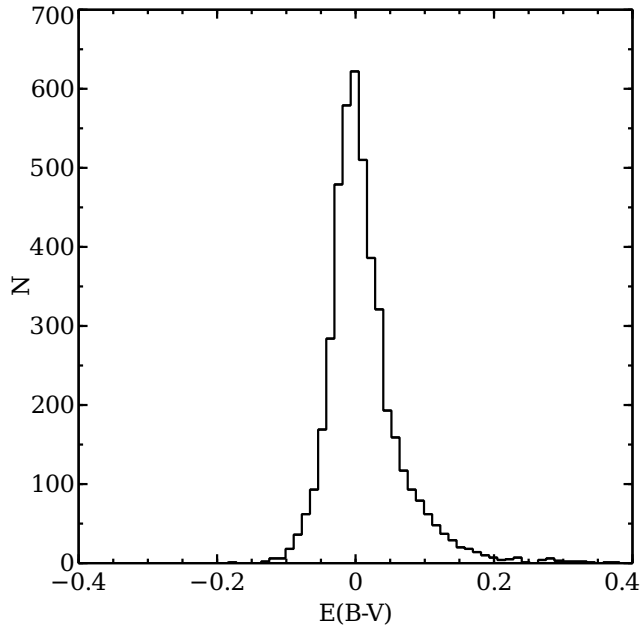}
\caption{The $E(B-V)$ distribution (derived by fitting each absorber's 
flux ratio with an SMC extinction law) for \uncertnum \ quasar spectra,
with no identified absorbers, from our quasar sample (Section
\ref{mgsample}). The standard deviation of $\sigma_{E(B-V)}$=0.025
quantifies the uncertainty in an individual estimate of $E(B-V)$ due to
quasar-to-quasar SED variations (Section \ref{sub:Determining-E(B-V)}).}
\label{cap:uncertainty}
\end{figure}

\subsection{Uncertainty in the $\bmath{E(B-V)} $ 
determinations}\label{sub:Evaluating-the-method}
In order to test the accuracy of the method for calculating the
$E(B-V)$ for a particular absorber\footnote{A test of the method
  involved calculating $E(B-V)$ estimates for the sample of \caii
  absorbers studied by \citet{2005MNRAS.361L..30W} and no systematic
  differences with either redshift or $E(B-V)$ were found.}, we
obtained estimates for the amounts of dust in a subsample of
\uncertnum \ quasar spectra with no identified absorbers from our
quasar sample (Fig. \ref{cap:uncertainty}). For the non-absorber
quasar sample we create a simulated set of absorbers with
$z_{\mathrm{abs}}$ and $z_{\mathrm{abs}}/z_{\mathrm{qso}}$
distributions which are consistent with the corresponding redshift
distributions in the absorber sample. We then obtain $E(B-V)$
estimates for this sample using the prescription in Section
\ref{sub:Determining-E(B-V)}. The distribution is centred on
$E(B-V)=0.0$ indicating no detectable systematic bias in the
determination of $E(B-V)$, with an encouragingly small uncertainty,
$\sigma_{E(B-V)} \sim 0.025$, in the $E(B-V)$ estimates for individual
quasars\footnote{It is necessary to modify how the quasar control
  sample is defined when we approach the extremes of the $m_i$ plane
  where there are not enough quasars to make up the control-spectrum
  using the scheme described in Section \ref{sub:controlSpectrum}. In
  such a case we use the 100 quasars closest to the faintest/brightest
  quasar at a given redshift. The effectiveness of the modified sample
  definition is demonstrated by calculating $E(B-V)$ estimates for a
  sample of the 51st brightest quasars using control spectra defined
  using both the $\pm$50 neighbouring quasars and the next 100 fainter
  quasars. Average differences in the $E(B-V)$ estimates of only 0.001
  at the faint end and 0.0025 at the bright end, demonstrate that any
  systematic bias in the $E(B-V)$ estimates is insignificant.}. The
spread arises from intrinsic quasar-to-quasar SED variations.

\begin{figure}
\includegraphics[width=\columnwidth]{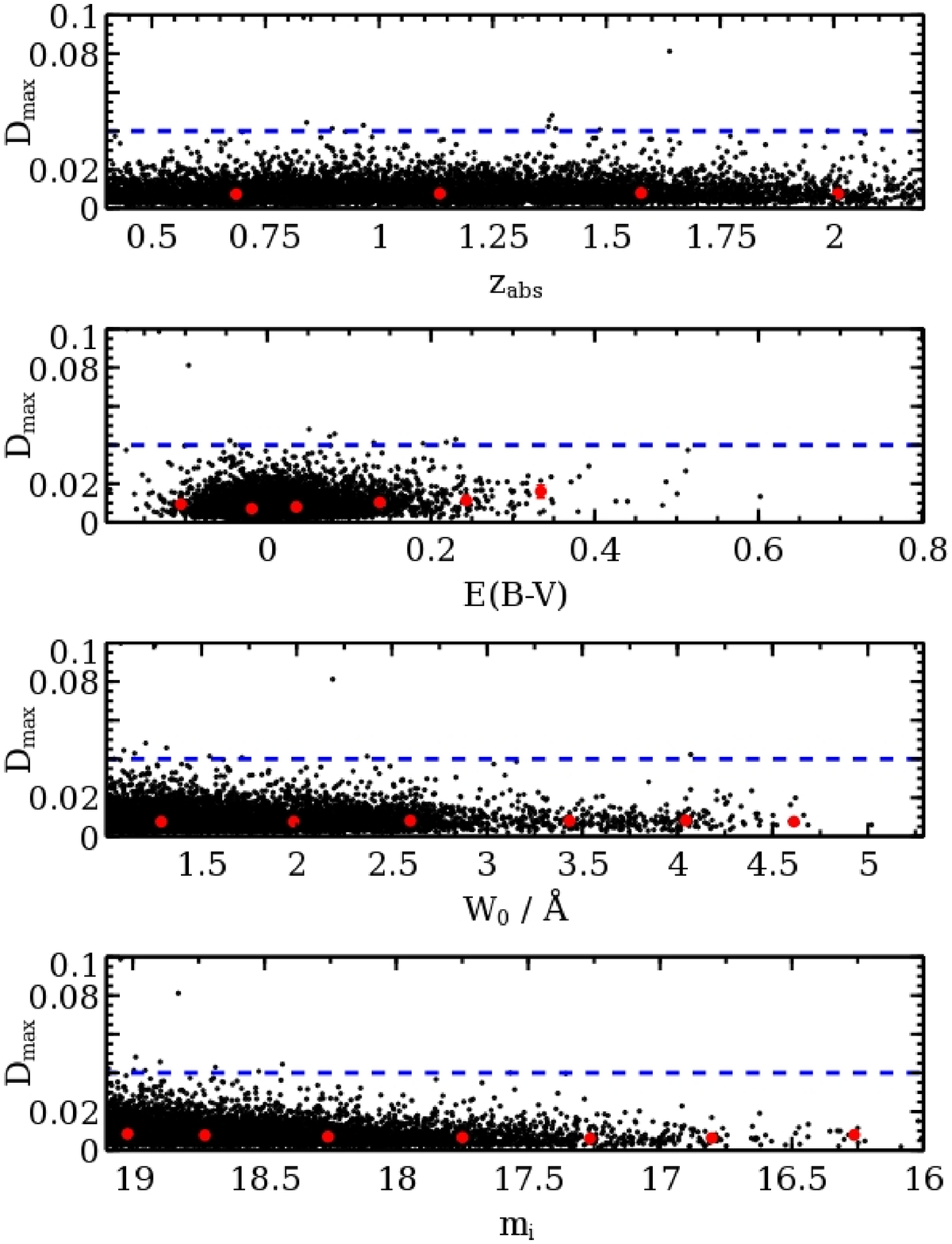}
\caption{The Kolmogorov-Smirnoff-type $D_{\mathrm{max}}$ statistic for
the base sample of \mgcap absorbers, fitted with an SMC-type
extinction law, as a function of (top to bottom panels) absorber
redshift, $E(B-V)$, equivalent width and quasar magnitude. The median
points show no significant bias in the quality of fit as a function of
any of the parameters. The horizontal line indicates the
$D_{\mathrm{max}}$=0.04 threshold, used to define systems with a
`poor fit' to the SMC extinction law.}
\label{cap:ks}
\end{figure}

Notwithstanding the use of an individual control spectrum for each
quasar, quasars with intrinsically unusual SEDs lead to flux ratio
spectra still dominated by the variation among quasar SEDs. The
presence of such variations in turn leads to poor determinations of
$E(B-V)$. We have estimated the quality of fit of each absorber's flux
ratio with an extinction curve by using a Kolmogorov-Smirnoff-like
statistic. Our test uses the cumulative difference between the
normalised best-fit extinction curve and normalised absorber's flux
ratio, to identify a systematic shape difference and thus
poor-fit. The maximum cumulative difference, or $D_{\mathrm{max}}$, is
obtained for each absorber's SMC, LMC and MW extinction curve fit.
The $D_{\mathrm{max}}$ distributions appear to show no evidence of
significant bias as a function of absorber redshift, $E(B-V)$, EW and
quasar magnitude (Fig. \ref{cap:ks}).

Visual inspection of quasar spectra with large values of
$D_{\mathrm{max}}$ reveals a mixture of quasars showing i) strong
\fetwo~ emission, ii) very blue continua or iii) red, but 'flat',
spectra without the curvature expected due to conventional extinction
curves. The number of such quasars is small, consistent with the tails
of the distribution evident in Fig.~\ref{cap:uncertainty}, due to the
small fraction of quasars with unusual intrinsic SEDs (unrelated to
any signature due to intervening absorbers). 

The 2$\sigma$ uncertainty in $E(B-V)$ resulting from quasar-to-quasar
SED variations corresponds to an $E(B-V)$ difference of 0.050 mag
(Fig. \ref{cap:uncertainty}). Such a difference produces a value of
$D_{\mathrm{max}}$=0.04 and we therefore define spectrum flux ratio
fits with $D_{\mathrm{max}}$$>$0.04 as possessing a poor fit to the
SMC extinction curve. The number of such spectra (17/\basenum)
represent an extremely small fraction of the total (0.2 per cent). We
exclude such objects from further analysis until we consider the
detailed shape of the \mg \ absorber extinction curves in
Section~\ref{dusttype}.

The effectiveness of the $D_{\mathrm{max}}$ statistic is verified by
noting that the flux ratio spectra with multiple ($>2$) absorption
systems exhibit systematically larger values of $D_{\mathrm{max}}$
than spectra with  single or double absorbers. The differences are due to the
presence of multiple extinction signatures at different redshifts,
providing justification for the use of single and double absorber lines of sight
only (Section \ref{mgsample}). 

\section{Dust obscuration bias}\label{sec:dustobsbias}
\begin{figure*}
\includegraphics[width=0.4\textwidth]{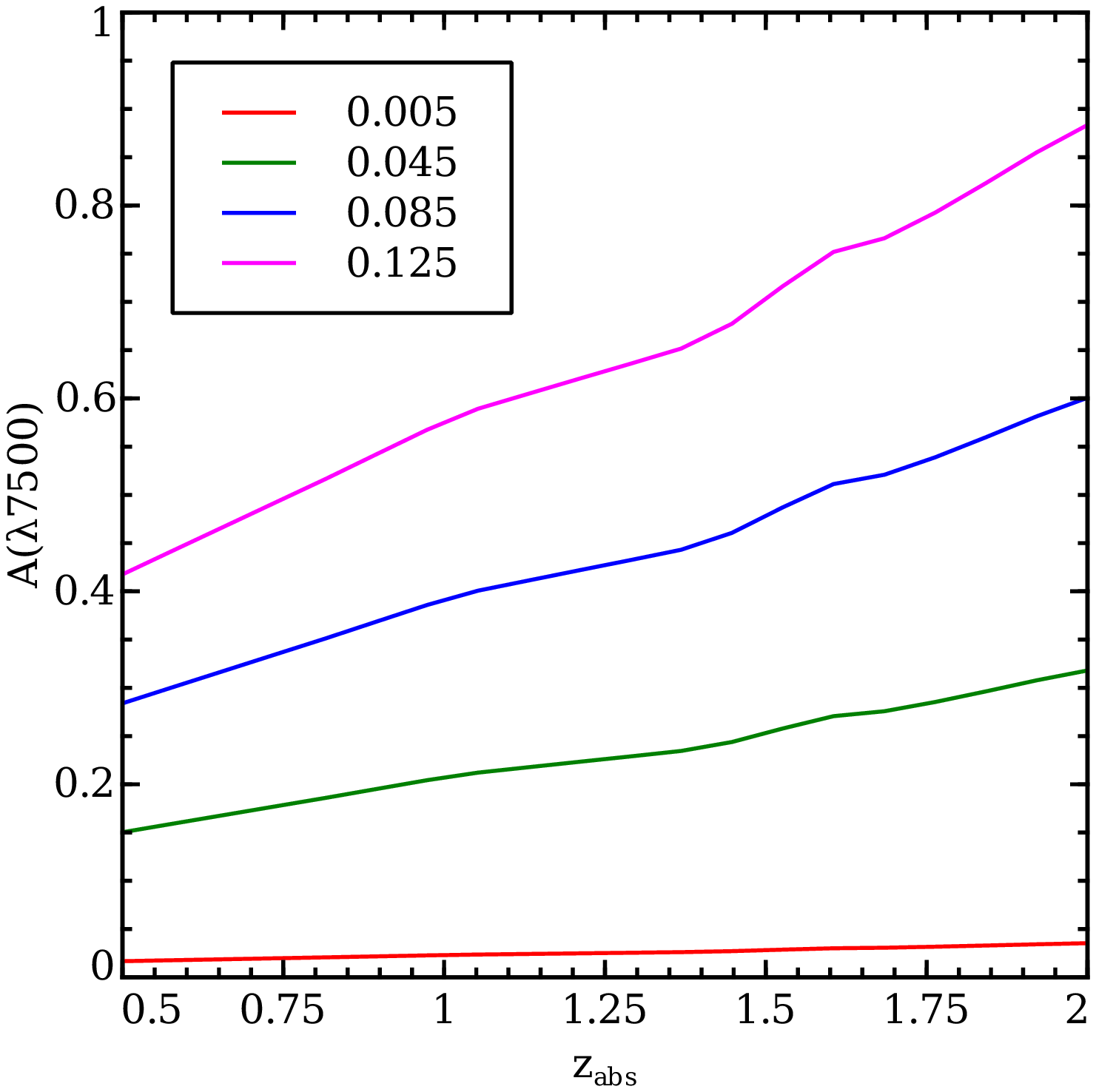}
\hspace{5mm}
\includegraphics[width=0.4\textwidth]{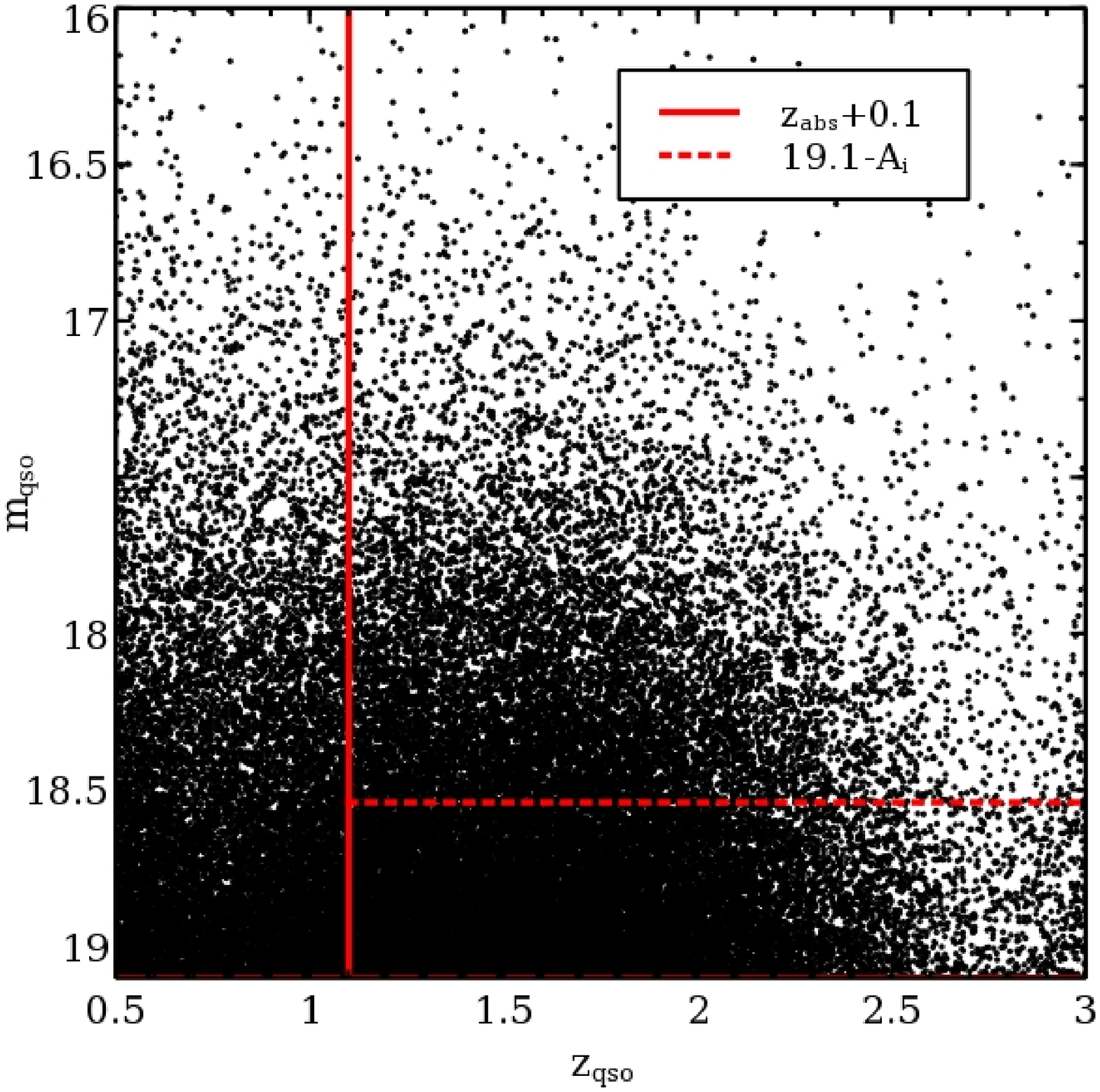}
\caption{Left-hand panel: the observed ($i$-band) extinction caused by
intervening LMC-type dust as a function of absorber redshift. The
different curves are for different values of $E(B-V)$. Right-hand
panel: the $i$-band magnitude versus quasar redshift for all objects
in the quasar sample (Section \ref{mgsample}). Objects to the
right of the vertical red line are quasars which are $\Delta z$$>$0.1
more distant than an absorber (at $z_{abs}=1.0$), and the
horizontal dashed line corresponds to the detection limit of quasars
placed behind the absorber with $E(B-V)=0.125$. The extinction,
($A_{i}$), is calculated as illustrated in the left-hand
panel.}\label{cap:biascalc}
\end{figure*}

Extinction of quasar light due to the presence of dust in intervening
absorbers, leads to objects being lost from magnitude-limited quasar
surveys \citep{1993ApJ...402..479F, 2006MNRAS.367..211W}. To
illustrate the effect of extinction by intervening dust on our sample,
the left-hand panel of Fig. \ref{cap:biascalc} shows the observed
extinction at 7500\ang (the effective wavelength of the SDSS
$i$-band). Additionally, extinction due to an intervening absorber
leads to a degredation of the spectrum S/N and hence additional
absorbers are lost from the sample, as their S/N falls below the
$7\sigma$ \mg \ absorber detection threshold (Section \ref{mgsample}).

\begin{figure}
\includegraphics[width=\columnwidth]{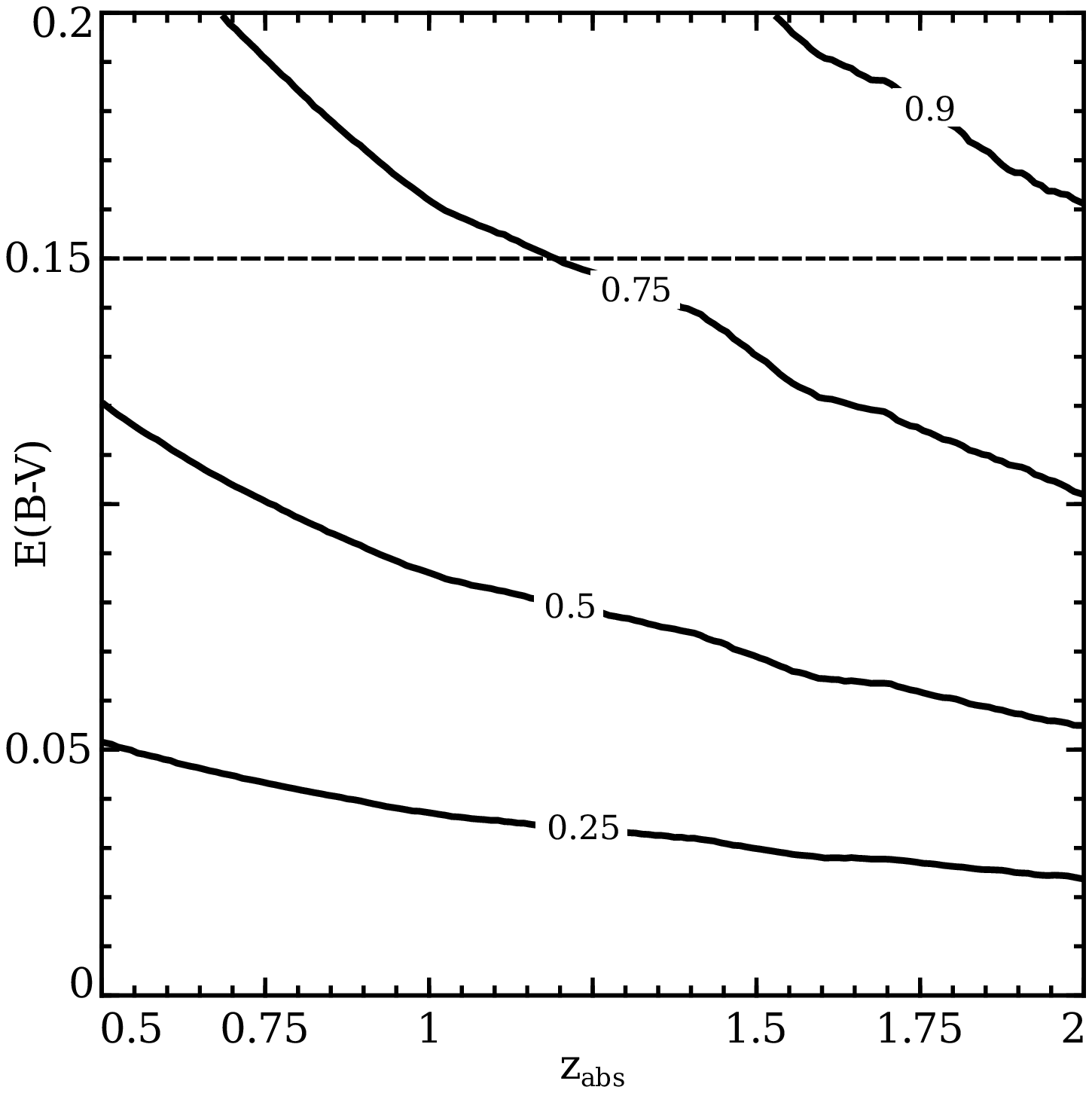}
\caption{Left-hand panel: Dust obscuration bias for SDSS quasar
  absorption systems as a function of absorber redshift ($z_{abs}$)
  and $E(B-V)$. The contours represent the probability that an
  individual absorber will \emph{not} be observed. The dashed
  horizontal line represents the $E(B-V)=0.15$ value which corresponds
  to a dust obscuration bias of 0.85 at $z_{\mathrm{abs}}=2$. In
  Section \ref{sec:redevol} we use this $E(B-V)$ value to define a
  conservative $E(B-V)$-cut to examine any statistical trends with
  redshift }
\label{cap:surfbias}
\end{figure}

\subsection{Extinction effect}\label{sub:bias-ext}
We have used the magnitude and redshift distribution of the quasars in
the quasar sample (Section \ref{mgsample}) to obtain a probability
that a quasar behind a dusty absorber will be lost from the
sample. The presence of an absorber in a quasar spectrum leads to an
extinction of quasar light $A_i$ (Fig.~\ref{cap:biascalc}, left-hand
panel). If the quasars in the sample\footnote{We consider only
  absorbers that are displaced at least $\Delta z$$>$0.1 from the
  target quasar} were placed behind the absorber, objects fainter than
$19.1-A_i$ will be lost below the $m_i=19.1$ flux limit. We can
therefore quantify the probability an absorber of specified $E(B-V)$
and redshift is lost from the sample via

\begin{equation} 
P_{\mathrm{dob}}=1-\frac{N_{\mathrm{qso}}\left(\leq19.1-A_{\mathrm{i}}\mathrm{,}\;z_{\mathrm{abs}}<z_{\mathrm{qso}}-0.1 \right)}{N_{\mathrm{qso}}\left(\leq19.1\mathrm{,}\;z_{\mathrm{abs}}<z_{\mathrm{qso}}-0.1 \right)},\label{eq:plossdob}
\end{equation}

\noindent where $N_\mathrm{qso}$ is the number of quasars satisfying
the condition in brackets. An illustration of the bias calculation is
shown in the right-hand panel of Fig.~\ref{cap:biascalc} for an
absorber at $z_{\mathrm{abs}}=1.0$ with $E(B-V)=0.125$.

\subsection{Signal-to-noise ratio effect}\label{sub:bias-snr}
For absorbers that remain within the sample there is a further
reduction in the probability of detection due to the decrease in the
S/N of the quasar spectrum as the quasar becomes fainter because of
the extinction from dust. Taking only the quasars that remain
above the m=19.1-$A_{i}$ threshold determined above, we can determine
how a quasar spectrum's S/N degrades for a given extinction $A_i$ by
examining the median S/N versus $m_i$ for the quasar spectra. Consider
the effect of placing an absorber of specified $E(B-V)$ and redshift
in front of a quasar with a given $m_i$, and hence S/N. The extinction
due to the absorber makes the quasar fainter and the \mg \ absorber
may no longer be detectable given the reduced S/N of the spectrum. An
\mg \ absorber will be lost from the sample when
  
\begin{equation} 
\mathrm{S/N}_{\mathrm{abs}}\times\frac{\mathrm{S/N}_{\mathrm{qsoext}}}{\mathrm{S/N}_{\mathrm{qsoabs}}}\leq7\sigma\label{eq:condmsnqsoext}
\end{equation}

\noindent holds true, where $\mathrm{S/N}_{\mathrm{abs}}$ is the S/N
of the \mg \ absorption feature, $\mathrm{S/N}_{\mathrm{qsoext}}$ is the
degraded S/N of the quasar spectrum, $\mathrm{S/N}_{\mathrm{qsoabs}}$
is the S/N of the absorber's target quasar spectrum, and 7$\sigma$ is
the \mg \ absorber detection threshold. We can therefore obtain a
separate probability that an absorber will be lost from the sample due
to the reduction in the spectrum S/N, 

\begin{equation} 
  P_{\mathrm{S/N}}=1-\frac{N_{\mathrm{qso}}\left(\mathrm{S/N}_{\mathrm{abs}}\times\frac{\mathrm{S/N}_{\mathrm{qsoext}}}{\mathrm{S/N}_{\mathrm{qso}}}\leq7\sigma\right)}{N_{\mathrm{qso}}\left(\mathrm{S/N}_{\mathrm{abs}}\times\frac{\mathrm{S/N}_{\mathrm{qsoext}}}{\mathrm{S/N}_{\mathrm{qsoabs}}}\geq7\sigma\right)},\label{eq:plosssnr}
\end{equation}

\noindent where $\mathrm{S/N}_{\mathrm{qso}}$ is the S/N of a given
quasar spectrum in the base quasar sample. $P_{\mathrm{S/N}}$ is
calulated using only the quasars that remain in the sample following
the determination of $P_{\mathrm{dob}}$ in Section \ref{sub:bias-ext}.

\subsection{Correcting the statistics}\label{sub:corstat}
The base absorber sample is highly incomplete at large values of
$E(B-V)$, and we provide a correction to the $E(B-V)$-distributions
for both the extinction and S/N effects described in this
section. Each absorber is assigned an $E(B-V)$-weighting due to both
obscuration effects according to

\begin{equation}
w=\frac{1}{1-P_{\mathrm{dob}}\left(E(B-V),z\right)}\times\frac{1}{1-P_{\mathrm{S/N}}\left(E(B-V),z\right)},\label{eq:weightebv_old}
\end{equation}

\noindent which allows the observed $E(B-V)$-distributions to be
corrected. The weighting, $w$, becomes undesirably large for absorbers
with large $E(B-V)$, where the probability an absorber is included in
the sample is small. We therefore impose a conservative upper limit to
the value of $w$=6.67, equivalent to the probability an absorber is
included in the sample of 0.15. The multiplicative prescription for
calculating $w$ is valid as we only apply the S/N corrections to the
population of absorbers which are \emph{not} lost due to dust
extinction effects. Thus, suppose we have a sample of absorbers from
which 30 per cent of quasars fall below the magnitude limit due to
extinction. The reduction in the S/N is calculated for the remaining
70 per cent of quasars and (in the case that the S/N effect reduces
the number of absorbers by 10 per cent) the fractional completeness of
the absorber sample is $0.7 \times 0.9 = 0.63$.

\subsubsection{Accuracy of the corrected $E(B-V)$ estimates}\label{sec:acccors}

The observed $E(B-V)$ distributions show a spread at fixed EW due in
part to quasar-to-quasar SED-variations, quantified in Section~3.3 and
illustrated in Fig.~\ref{cap:uncertainty}. In principle, a full
deconvolution of the SED-induced variations could be undertaken using
the observed $E(B-V)$ distributions to recover the intrinsic $E(B-V)$
distributions. However, such a procedure is both involved and unstable
in portions of the $E(B-V)$ {\it versus} EW plane.

The use, in Section~\ref{analyres}, of the median $E(B-V)$ values for
the absorber population as a function of absorber EW already mitigates
the impact of the intrinsic quasar SED-induced variations. However,
the procedure described in the previous sub-sections does result in a
small overestimation of the population median $E(B-V)$. The worst-case
situation occurs when considering a population of absorbers with no
dust, subject to the quasar SED-induced spread. A histogram of
`observed $E(B-V)$', exactly as shown in Fig.~\ref{cap:uncertainty} is
thus obtained. The observed median $E(B-V)$ is only a few thousandths
of a magnitude from zero but, following correction of the $E(B-V)$
distribution according to Equation \ref{eq:weightebv_old}, the median
$E(B-V)$ becomes +0.016 mag.  The origin of the positive value derives
in part to the small asymmetry to positive values in the histogram but
(mostly) to the application of the correction scheme only to objects
with apparently positive $E(B-V)$ values.

The overestimation of the median $E(B-V)$ is small under worst-case
conditions and becomes much smaller\footnote{The offset is $\la$+0.005
  for the median $E(B-V)$ estimate of the absorber sample with \mg
  EW $>3$\,\AA.} as populations with dust-induced $E(B-V)$ are
considered and the fraction of absorbers scattered artificially to
negative $E(B-V)$ drops. Our conclusions in the paper, which derive
from the presence of median $E(B-V)$ values $\ga0.07$\,mag for
absorbers with high \mg EW, are essentially unaffected by the form
of correction adopted to account for the small overestimation of the
median $E(B-V)$ values. Results obtained after applying a uniform
correction of -0.016\,mag to all the median estimates\footnote{This
  correction is performed by subtracting 0.016 from each $E(B-V)$
  value in the absorber sample and recalculating the weighted median.}
and also no correction at all are not significantly different. We thus
adopt a simple scheme that closely approximates the actual situation,
applying a correction of -0.016 mag to the median $E(B-V)$ estimates
of the absorber samples with \mg EW $\leq3.0$\,\AA \ and no
correction to the median $E(B-V)$ estimates for \mg EW $>3.0$\,\AA.

\begin{figure}
\includegraphics[width=\columnwidth]{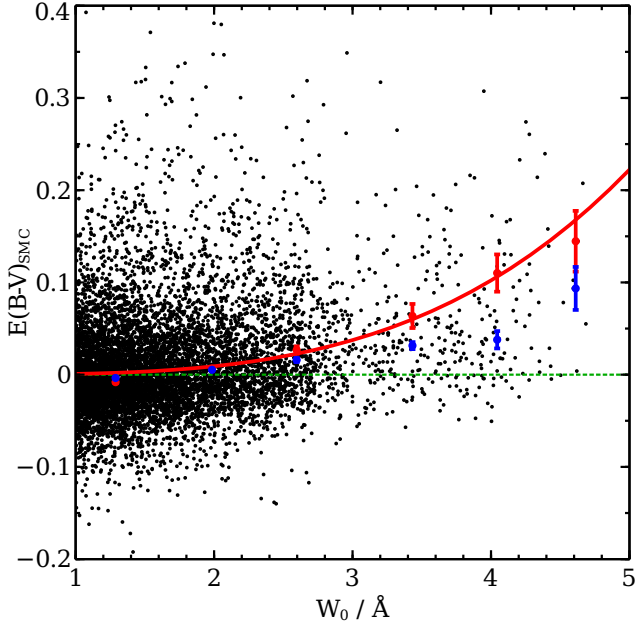}
\caption{$E(B-V)_{\mathrm{SMC}}$ as a function of $W_0^{\lambda2796}$
  for a sample of \mgcap absorption systems with accurate
  $E(B-V)_{\mathrm{SMC}}$ values (i.e. those with
  $D_\mathrm{max}^{\mathrm{smc}}<0.04$). The blue points represent the
  median values for a constant equivalent width bin of 0.7\,\AA, and
  the red points represent the median values adjusted for dust
  obscuration bias. The error bars are calculated by using the
  `bootstrap' method, and the solid red line represents the best
  fitting power-law to the bias-corrected data. At low values of
  $W_{0}^{2796}$ there is little evidence for significant $E(B-V)$
  associated with the absorbers and the distribution of $E(B-V)$ values
  is dominated by the quasar SED-induced spread illustrated in
  Fig. 3. The true uncertainty in the estimation of the median $E(B-V)$ at low
  values of $W_{0}^{2796}$ is limited by systematics at the level of
  $\sim0.005$.}
\label{cap:fullebvew}
\end{figure}

\section{Results}\label{analyres}

The amount of dust in \mg \ absorption systems is generally found to
increase as a function of the \mg \ EW \citep{2006MNRAS.367..945Y,
2008MNRAS.385.1053M}. This increase has important implications for the
understanding of galaxy evolution, as EW for the saturated \mg \ 
absorbers measures the velocity spread in the gas
\citep{2006MNRAS.367..945Y} and there is a strong dependence of EW on
the level of associated star-formation activity
\citep{2009arXiv0912.3263M}.

\subsection{Dust and absorber Equivalent width}\label{sub:ewdep}

The accuracy of the $E(B-V)_{\mathrm{SMC}}$\footnote{To make the form
  of the extinction curve used to obtain $E(B-V)$ values explicit,
  where relevant, we use a subscript `LMC', `MW', `SMC',... in the
  remainder of the paper} estimate for each absorber was ensured by
removing all spectra with a poor fit to the SMC extinction curve,
i.e. $D_{\mathrm{max}}^{\mathrm{smc}}>0.04$ (Section
\ref{sub:Evaluating-the-method}). We chose the SMC curve as the
\emph{average} reddening properties of \mg \ absorbers are found to be
well characterised by an SMC-type extinction law (see Section
\ref{dusttype}).
 
\begin{figure*}
\includegraphics[width=0.7\textwidth]{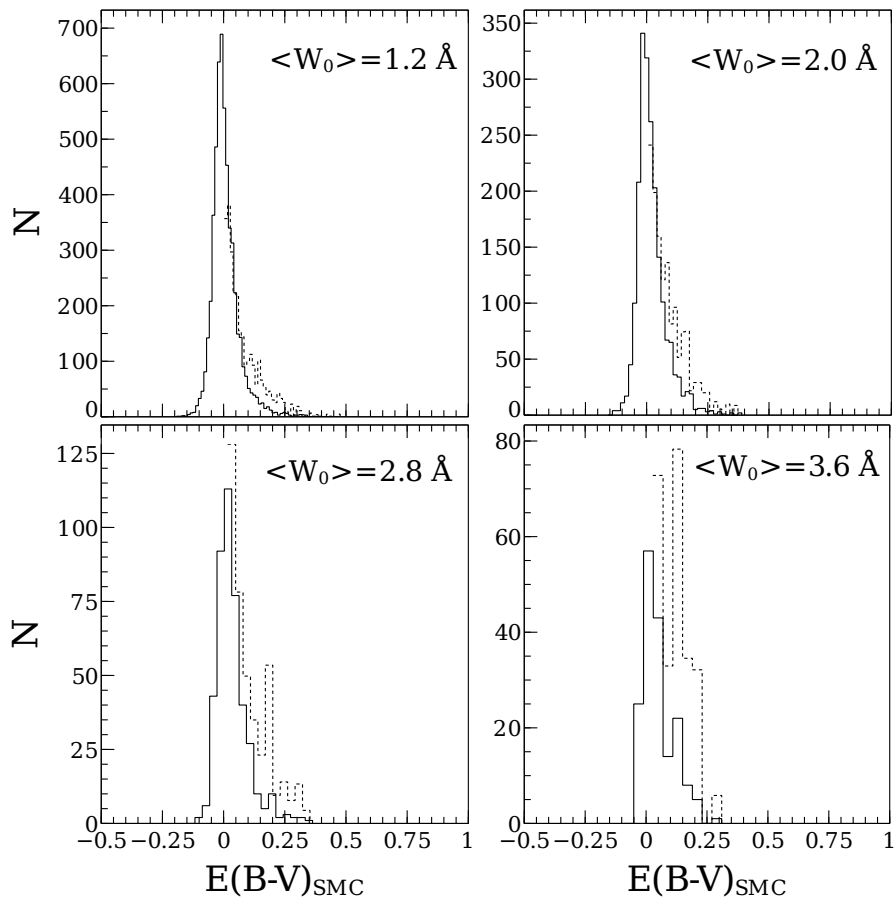}
\caption{The $E(B-V)_{\mathrm{SMC}}$ distributions of \mgcap
  absorption systems for different equivalent width ranges. The solid
  line represents the distribution of observed
  $E(B-V)_{\mathrm{SMC}}$-estimates, and the dotted line corresponds
  to the distribution corrected for bias (Section
  \ref{sec:dustobsbias}). The dotted lines represent estimates of the
  intrinsic $E(B-V)_{\mathrm{SMC}}$ distribution, which exhibit more
  pronounced high-$E(B-V)$ tails.}
\label{cap:ewbindist}
\end{figure*}

\subsubsection{$E(B-V)$ Distributions}
The dependence of dust content on \mg \ EW is shown in
Fig. \ref{cap:fullebvew}. The median $E(B-V)_{\mathrm{SMC}}$ is
calculated for absorbers with EWs in the intervals 1.0-1.7\,\AA,
1.7-2.4\,\AA, 2.4-3.1\,\AA, 3.1-3.8\,\AA, 3.8-4.5\,\AA, and
4.5-5.0\,\AA. The equivalent median $E(B-V)_{\mathrm{SMC}}$ values
incorporating the bias corrections in Equation \ref{cap:fullebvew} are
also shown. The error bars have been calculated using the `bootstrap'
method but, at low $W_{0}^{2796}$, the errors are limited by
systematic uncertainties at the level of $\sim0.005$. The line in
Fig. \ref{cap:fullebvew} corresponds to the best fitting power-law to
the median bias-corrected $E(B-V)_{\mathrm{SMC}}$ values of the form

\begin{equation}
\left\langle E(B-V)_{\mathrm{SMC}}\right\rangle =A\left(W_{0}^{\lambda 2796}\right)^{\alpha},\label{eq:fullebvew}
\end{equation}

\noindent where $A=\plnorm$ and $\alpha=\plpower$ and the model is
applicable for 1.0$\le$$W_0$$\le5.0$\,\AA.

The number and distribution of objects evident in
Fig.~\ref{cap:fullebvew} with $E(B-V)_{\mathrm{SMC}}<0$ are consistent with the
expected observational uncertainty in the $E(B-V)$ determinations
(Section \ref{sub:Evaluating-the-method}). A series of $E(B-V)_{\mathrm{SMC}}$
distributions for different bins of mean \mg \ EW 1.2\,\AA, 2.0\,\AA,
2.8\ang and 3.6\,\AA, are shown in Fig.~\ref{cap:ewbindist}.

\subsubsection{Comparison with other work}
\citet{2008MNRAS.385.1053M} provide a parameterisation of the $E(B-V)$
content of strong \mg \ absorbers as a function of EW that has started
to see wide use in the literature (e.g. \citealp{2009ApJ...697.1725E};
\citealp{2009ApJ...699...56S}). \citeauthor{2008MNRAS.385.1053M}'s
investigation presented \emph{observed-frame} $E(B-V)_{\mathrm{SMC}}$
distributions, which contrasts with the absorber rest-frame
parametrizations, made possible by the direct use of the SDSS quasar
spectra, presented here.

The observed-frame $E(B-V)_{\mathrm{SMC}}$ dependence on \mg
\ absorber EW (equation \ref{eq:fullebvew} of
\citeauthor{2008MNRAS.385.1053M}) can be related to our rest-frame
$E(B-V)_{\mathrm{SMC}}$ dependence via

\begin{equation}
E(B-V)_{\mathrm{obs}}=E(B-V)_{\mathrm{rest}}\left ( 1+z \right )^{1.2}.
\label{obsebv}
\end{equation}

\begin{figure}
\includegraphics[width=\columnwidth]{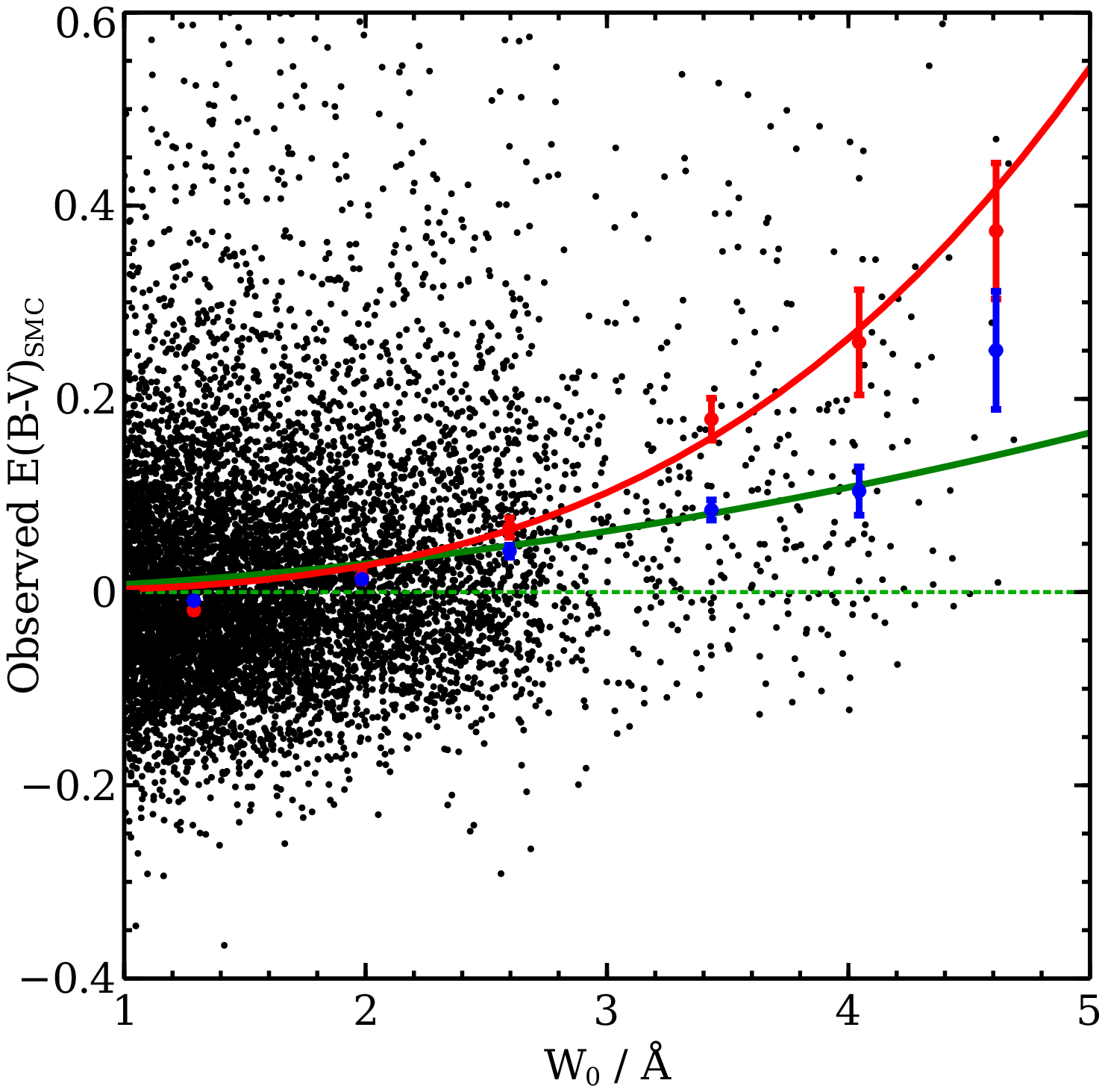}
\caption{The \emph{observed-frame}
  $E(B-V)_{\mathrm{SMC}}$-distribution of \mg \ absorbers as a
  function of $W_0^{\lambda2796}$. The observed-frame
  $E(B-V)_{\mathrm{SMC}}$ values (calculated using Equation
  \ref{obsebv}) are plotted to enable comparison with the
  \citet{2008MNRAS.385.1053M} result shown as the solid-green
  line. The blue points are the median observed-frame
  $E(B-V)_{\mathrm{SMC}}$ values calculated using EW bins of
  $\Delta$EW=0.7\,\AA, and the red points represent the median values
  adjusted for dust-obscuration bias. The error bars are calculated by
  using the 'bootstrap' method, and the red line represents the best
  fitting power-law to the bias-corrected data. The true uncertainty
  in the estimation of the median E(B-V) at low values of
  $W_{0}^{2796}$ is limited by systematics at the level of
  $\sim0.005$.}
\label{cap:compwork}
\end{figure}

Fig.~\ref{cap:compwork} shows our data in the $E(B-V)_{\mathrm{obs}}$
versus EW plane (c.f. fig.~9 of \citeauthor{2008MNRAS.385.1053M}))
along with the median $E(B-V)_{\mathrm{obs}}$ points, median
dust-corrected $E(B-V)_{\mathrm{SMC}}$ values and the power-law fit
(Equation \ref{eq:fullebvew}), all transformed into the observed
frame. There is good agreement between the
\citeauthor{2008MNRAS.385.1053M} observed median
$E(B-V)_{\mathrm{SMC}}$ values and those determined here. However, our
median dust-corrected $E(B-V)_{\mathrm{SMC}}$ values lie significantly
above \citeauthor{2008MNRAS.385.1053M}'s observed values. The large
differences can be ascribed to the ability to quantify the effects of
$E(B-V)$-bias on the observed \mg \ sample, set out in Section
\ref{sec:dustobsbias}. These dust obscuration effects while linked to
EW, also become more pronounced at higher redshift.

\subsection{Redshift evolution of $\bmath{E(B-V)}$}\label{sec:redevol}

The ability to constrain how the dust content of \mg \ absorbers
varies with cosmic time, from 3.3 Gyr at $z=2.0$ to 8.6 Gyr at
$z=0.5$, when a dramatic reduction in in the overall star formation
rate density of the Universe \citep{2004ApJ...615..209H,
  2009arXiv0912.3263M} occured, provides a significant constraint on
models of galaxy evolution.

Notwithstanding the quantitative corrections to the observed
$E(B-V)_{\mathrm{SMC}}$ distribution of the \mg \ absorbers (Sections
\ref{sec:dustobsbias} and \ref{sub:ewdep}) the sample suffers from
significant incompleteness for absorbers with large $E(B-V)$ at high
redshifts. To investigate the evolution of the absorbers as a function
of redshift we therefore confine our analysis to systems with
$E(B-V)_{\mathrm{SMC}}<0.15$. The adopted limit to
$E(B-V)_{\mathrm{SMC}}$ corresponds to that of an absorber 
$P_{\mathrm{dob}}$$>$0.85\footnote{Application of the bias-correction to
  absorbers with $P_{\mathrm{dob}}\ge0.85$ can cause undesirable divergence in
  the population statistics.} at the maximum absorber redshift,
$z$=2.0, employed (see Fig.~\ref{cap:surfbias}).

\subsubsection{Results and comparison with previous work}
The restricted absorber sample ($E(B-V)_{\mathrm{SMC}}<0.15$) was
divided into three redshift bins, with average redshifts of 0.77,
1.25, and 1.68. The resulting bias-corrected median rest-frame
$E(B-V)_{\mathrm{SMC}}$ are fit with a power-law of the form

\begin{equation}
E(B-V)\propto\left(1+z\right)^{\alpha},
\end{equation}

\noindent where $\alpha=-0.2\pm0.3$, which shows no significant evolution with redshift. This determination is
consistent with \citet{2008MNRAS.385.1053M} who find a modest
evolution of dust content with redshift ($\alpha=-1.1\pm0.4$).

\begin{figure}
\includegraphics[width=\columnwidth]{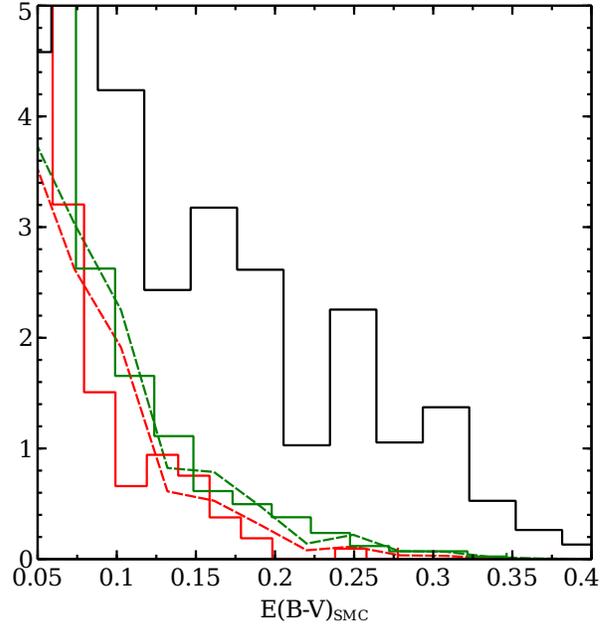}
\caption{The normalised
  distribution of absorbers for the full sample (in different redshift
  bins) as a function of $E(B-V)_{\mathrm{SMC}}$. The histograms have
  been normalised to enable comparison of redshift bins containing
  different numbers of absorbers. The black solid line corresponds to
  the distribution of absorbers in the low-redshift ($\langle z
  \rangle=0.77$) bin when corrected for dust bias effects. The green
  and red histogram are the observed distributions (i.e. uncorrected
  for dust bias) of $E(B-V)$ for the higher redshift bins ($\langle z
  \rangle=1.25$ and $\langle z \rangle=1.25$) respectively. The dotted
  lines represent the result of applying the dust obscuration effects
  to the black histogram at these higher redshifts. The consistency
  between the observed histograms and the predicted distributions
  again highlight that there is no evidence for significant evolution
  of the dust content of \mg \ absorbers as a function of redshift.}
\label{cap:fullebvz}
\end{figure}

While this analysis of the bias-corrected median points has
deliberately been confined to absorbers with
$E(B-V)_{\mathrm{SMC}}$$<$0.15 it is also possible to test whether the
entire observed absorber sample is consistent with no evolution as a
function of $z_{\mathrm{abs}}$. Taking the bias corrected
$E(B-V)_{\mathrm{SMC}}$ distribution of absorbers in the low-redshift
$\langle z_{\mathrm{abs}} \rangle =0.77$ slice we calculate the
observed $E(B-V)_{\mathrm{SMC}}$ distribution, by applying the inverse
of the bias corrections to the $\langle z_{\mathrm{abs}} \rangle=0.77$
intrinsic distribution, centred on the higher redshift slices $\langle
z_{\mathrm{abs}} \rangle=1.25$ and $\langle z_{\mathrm{abs}}
\rangle=1.68$ respectively (Fig.~\ref{cap:fullebvz}). The resulting
distributions for the higher redshift slices are consistent with the
observed histograms, confirming the lack of evidence for significant
evolution in the distribution of $E(B-V)_{\mathrm{SMC}}$ with
redshift.

\begin{figure}
\includegraphics[width=\columnwidth]{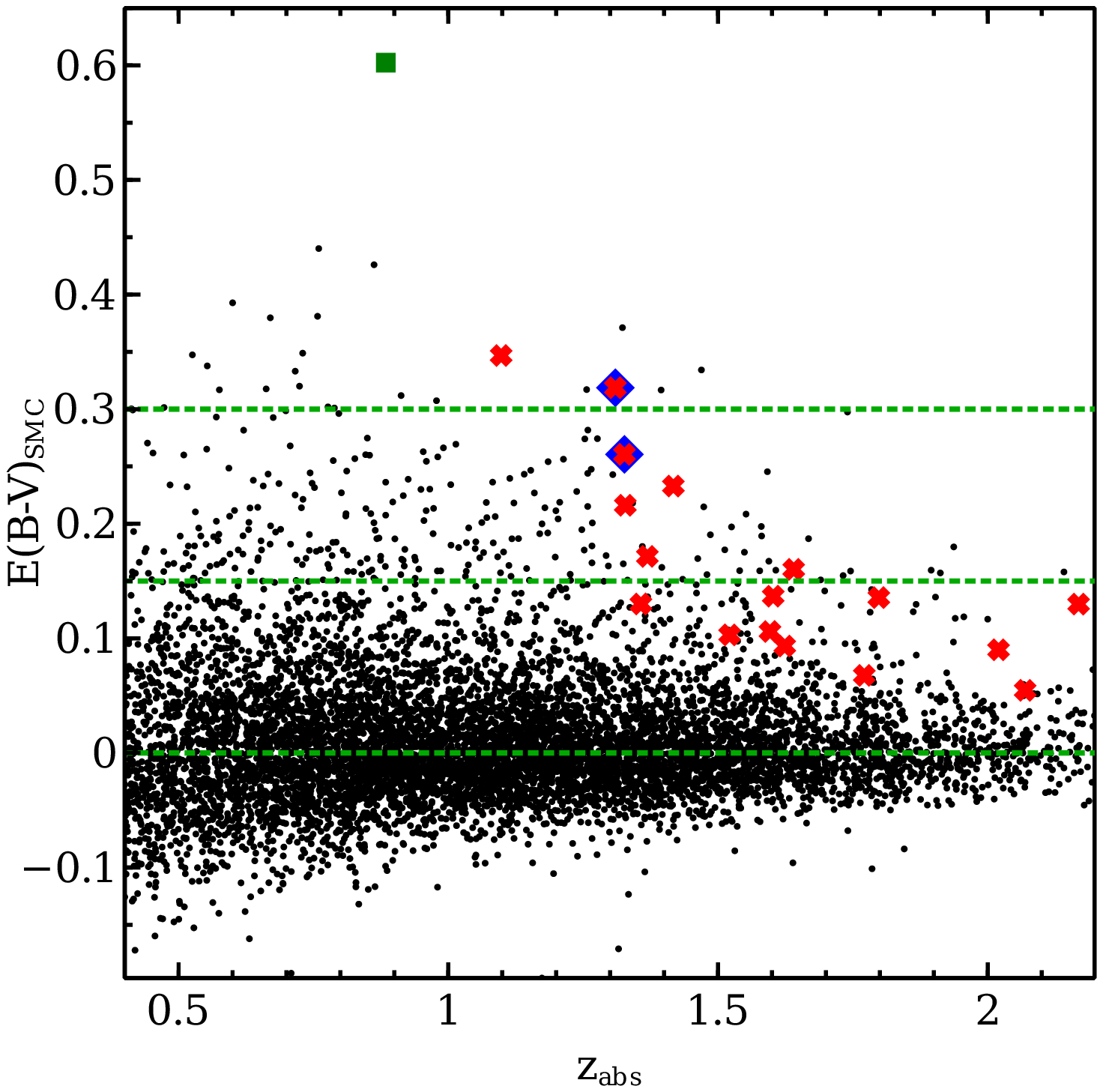}
\caption{The $E(B-V)_{\mathrm{SMC}}$ of \mgcap absorbers as a function
  of absorber redshift. The green square and blue diamonds correspond
  to the \citet{2010ApJ...708..742Z} and \citet{2008MNRAS.391L..69S}
  objects respectively. The red crosses represent the objects most
  likely to contain signatures of the 2175\ang bump, which make up
  the LMC coadd sample in Fig. \ref{cap:lmccoadd}. The horizontal line
  at $E(B-V)_{\mathrm{SMC}}=0.15$ is included to show which absorbers
  make it into the redshift evolution sample described in Section
  \ref{sec:redevol}, and the line at $E(B-V)_{\mathrm{SMC}}=0.30$
  shows which objects make it into the high-$E(B-V)$ sample described
  in Section \ref{sec:dustiest}.}
\label{cap:ebvz}
\end{figure}

\subsection{Nature of the dust}\label{dusttype}

Determining the nature of dust in \mg \ absorption line systems is
important as it provides a way to constrain the chemical evolution of
galaxies over a range of cosmic time. Of particular interest is
whether the strong $\sim2175$\ang feature observed in the spectrum of
the MW is present in our \mg \ absorber sample. In general, the SMC
extinction curve, rather than that of the MW, is found to best
describe the average reddening properties of \mg \ absorbers
\citep{2005pgqa.conf...86M, 2006MNRAS.367..945Y, 2008MNRAS.385.1053M}
with low $E(B-V)$. Taking advantage of the large sample of absorbers,
and specifically the availability of a number of absorbers
with significant $E(B-V)$, we investigated the form of the extinction
curve via construction of a composite spectrum and consideration of
the properties of the absorbers with the very highest estimates of
$E(B-V)$.

\subsubsection{An optimal high-$E(B-V)$ composite absorber spectrum}
Despite claims of being able to see evidence for the $2175$\ang bump
in single systems \citep{2008MNRAS.391L..69S}, individual
extinction curves have a low S/N. Using the much larger sample of
absorbers presented here, we can combine a large number of extinction
curves statistically, using an approach similar to
\citet{2005MNRAS.361L..30W}, to improve the S/N and increase the
sensitivity to the presence of any features.

An important consideration is the choice of which systems to combine.
The optimum number of spectra to co-add is a trade-off between the
number of spectra (in principle the more the better), and the median
$E(B-V)$ of the coadded spectra (which favours a smaller number of
spectra with the largest $E(B-V)$s).

\begin{figure}
\includegraphics[width=\columnwidth]{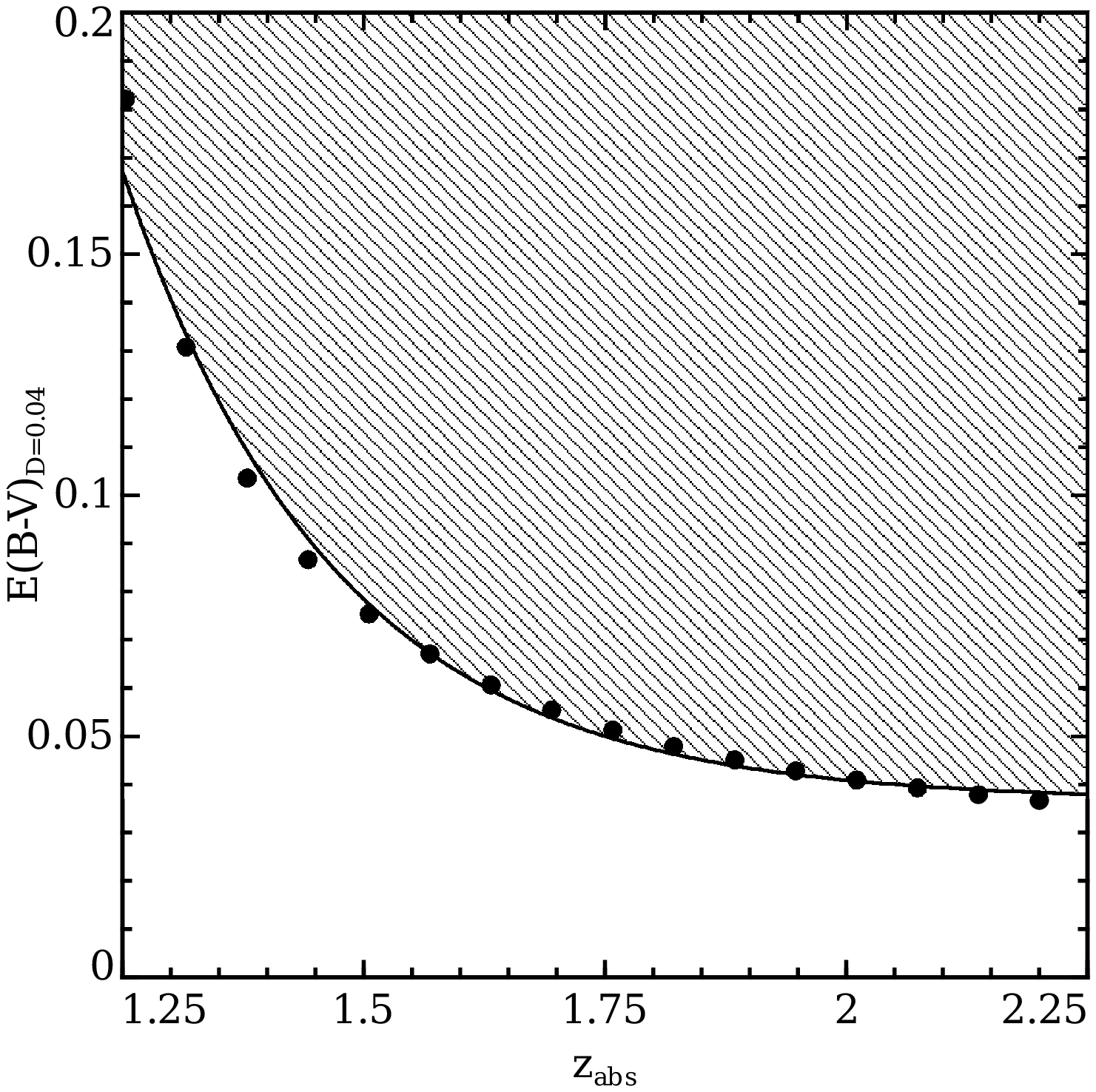}

\caption{The $E(B-V)_{\mathrm{SMC}}$ value which corresponds to a poor-fit
$D_{\mathrm{max}}=0.04$ statistic between an SMC and MW extinction
curve, as a function of redshift. The solid black line corresponds to
a parameterised fit to the black points, and the shaded region above
the line represents systems which can, in principle, be distinguished
as containing either SMC or MW type dusts on the basis of their SMC/MW
$D_{\mathrm{max}}$ statistic.}
\label{cap:natdustcomp1}
\end{figure}

We also need to consider the goodness-of-fit of the spectra that will
make up the composite spectrum, and so it was necessary to make a cut
based upon the $D_{\mathrm{max}}$ statistic (Section
\ref{dustquant}). However, it is conceivable that a system with a
large SMC-$D_{\mathrm{max}}$ value, and thus poor SMC fit, will in
fact be well characterised by a MW extinction curve. We certainly do
\emph{not} want to remove such objects which exhibit a MW type dust
signature from the composite sample. This $D_{\mathrm{max}}$
discrepancy (between SMC and MW extinction curves) becomes more
significant at higher $E(B-V)$ values, and we can quantify, for a
given absorber redshift, the $E(B-V)_{\mathrm{SMC}}$ value which
corresponds to a poor-fit $D_{\mathrm{max}}=0.04$. We have determined
how this $E(B-V)_{\mathrm{SMC}}$ cutoff varies as a function of
redshift and the results are shown in Fig \ref{cap:natdustcomp1}.

Absorption systems in the shaded region of Fig \ref{cap:natdustcomp1},
are those which satisfy the parameterised condition:

\begin{equation}
E(B-V)_{\mathrm{SMC}}>a e^{-bz}+c
\label{ebvcond}
\end{equation}

\noindent where $a=38.7$, $b=4.55$, and $c=0.04$. For such absorbers,
the form of the extinction curve, SMC or MW, can be determined via
their SMC/MW $D_{\mathrm{max}}$ statistics.

\begin{figure*}
\includegraphics[width=0.8\textwidth]{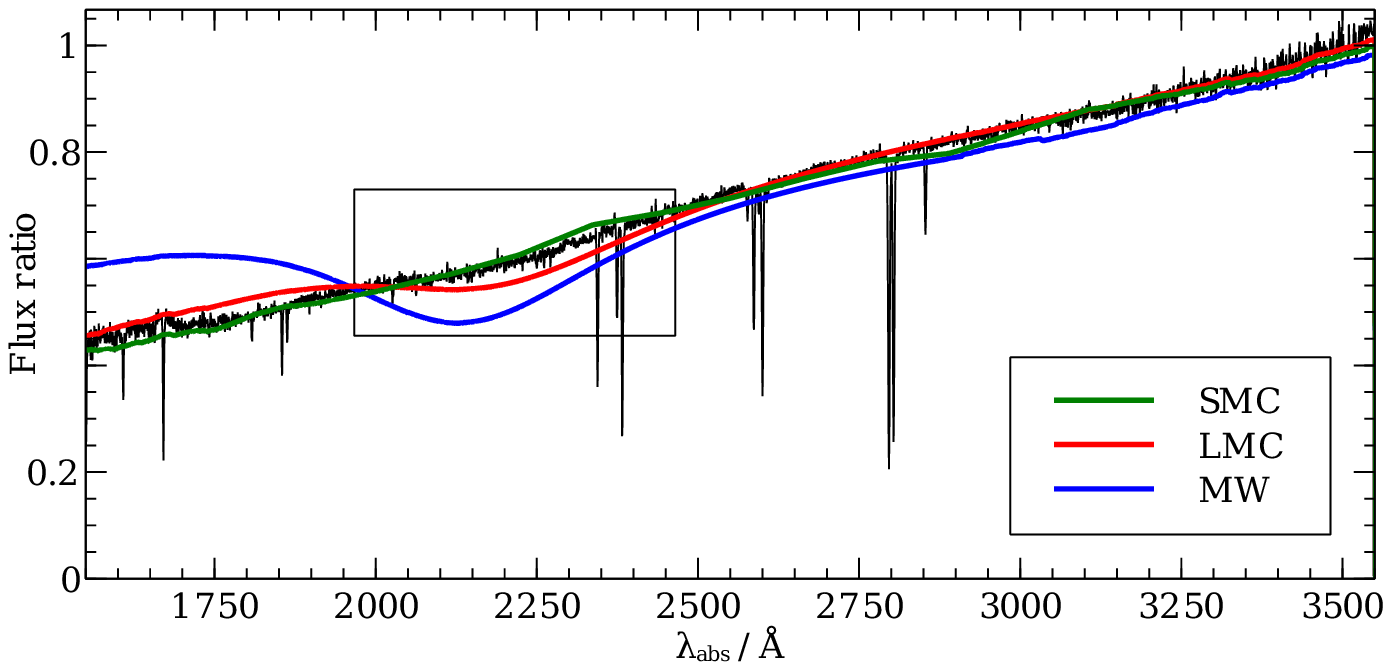}
\vspace{5mm}
\includegraphics[width=0.4\textwidth]{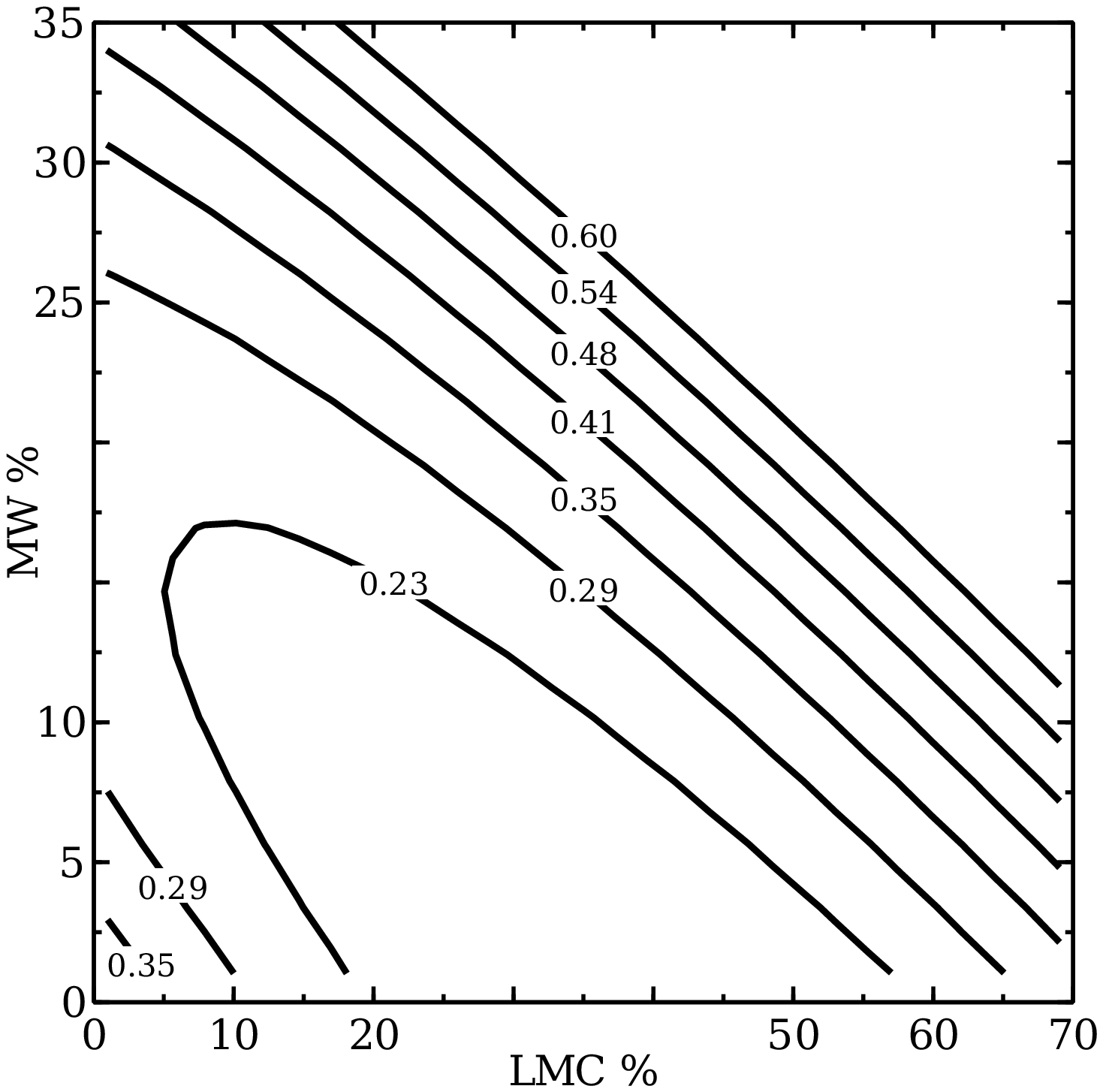}
\hspace{10mm}
\includegraphics[width=0.4\textwidth]{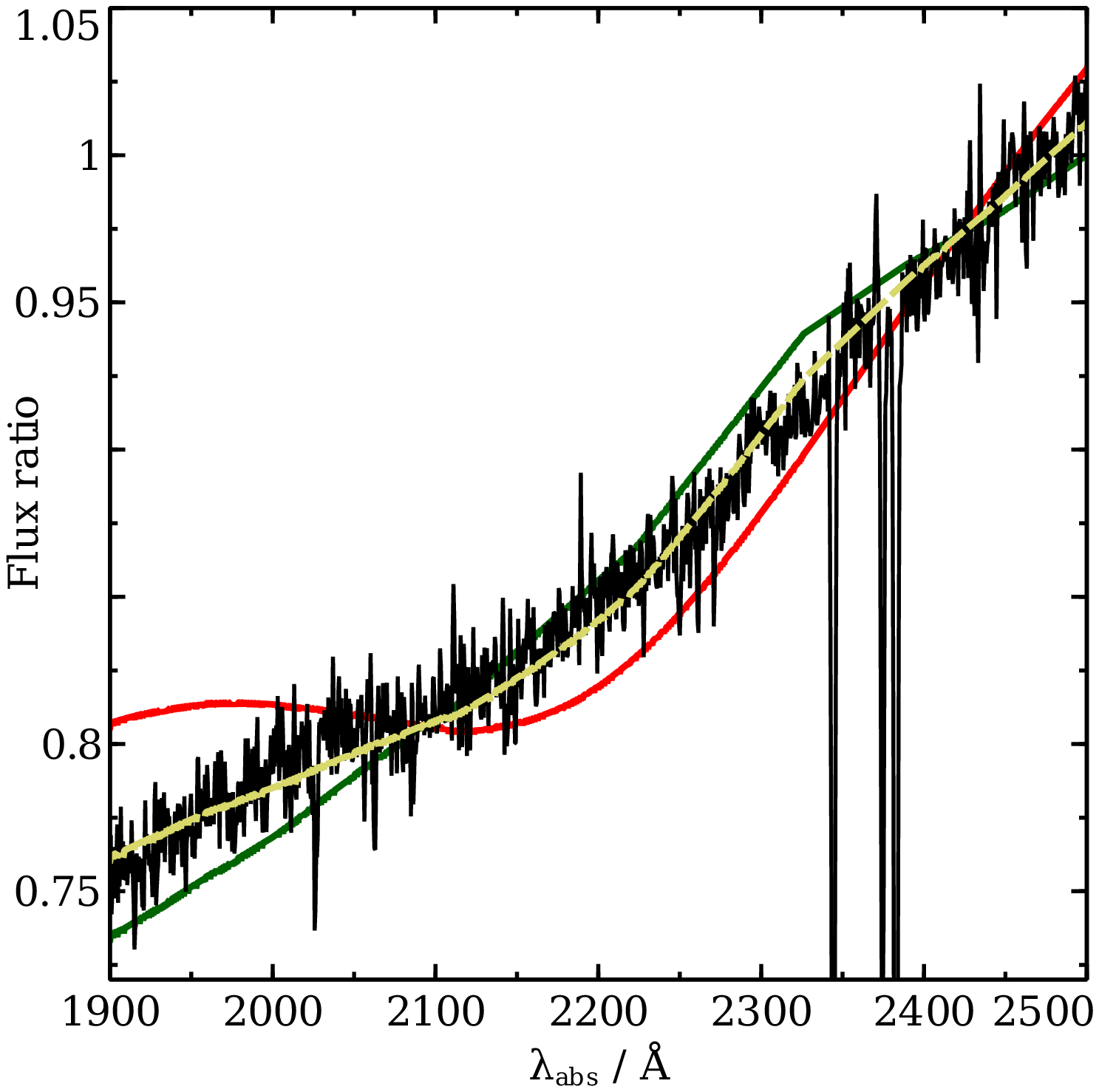}
\caption{Top panel: Wavelength versus flux ratio for a composite mean
  spectrum created from \compnum \ absorber flux ratios which satisfy
  Equation \ref{ebvcond}. Additional composites of each absorber's
  best fit SMC, LMC and MW extinction curves are also shown with mean
  $E(B-V)_{\mathrm{SMC}}=0.12$, $E(B-V)_{\mathrm{LMC}}=0.16$ and
  $E(B-V)_{\mathrm{MW}}=0.13$. The SMC curve is the best fit to the
  composite flux ratio spectrum. The black box defines the zoomed-in
  section around the 2175\ang bump. Bottom-left panel: The residuals
  between a mock coadded curve (containing $x$ per cent LMC spectra,
  $y$ per cent MW spectra and the rest SMC spectra), and the composite
  flux ratio. The best-fit corresponds to a composite containing only
  SMC and LMC extinction curves in the ratio 65:35. Bottom-right
  panel: The zoomed-in section of the composite spectrum with green,
  and red solid lines corresponding to the SMC and LMC extinction
  curve composites respectively. The gold-dashed line corresponds to
  the coadded curve containing SMC spectra and LMC in the ratio 65:35,
  which best characterises the composite flux ratio.}
\label{cap:coaddstrongmg}
\end{figure*}

\begin{figure}
\includegraphics[width=\columnwidth]{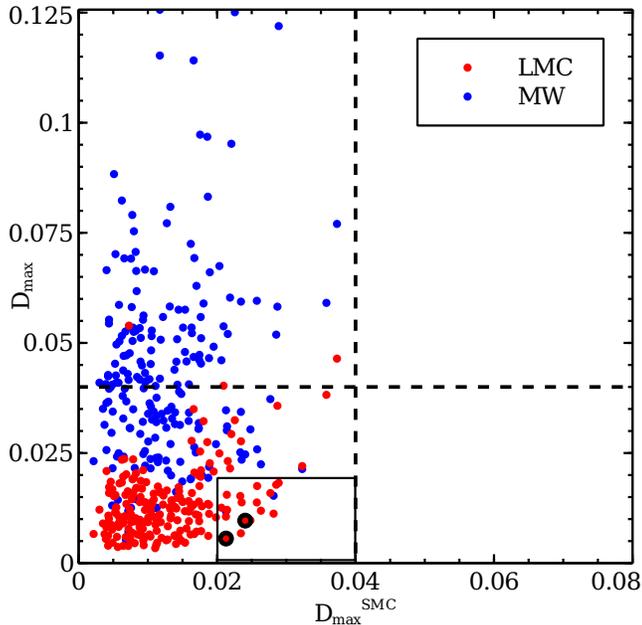}
\caption{$D_{\mathrm{max}}$ for the SMC curve versus the
  $D_{\mathrm{max}}$ for the LMC curve (\emph{red-points}) and the MW
  curve (\emph{blue-points}). The absorbers plotted are the 211 where
  discrimination between different extinction curves is possible
  according to the condition of Equation~\ref{ebvcond}. The dashed
  lines represent the average poor-fit cutoff (Section
  \ref{sec:dustobsbias}) of $D_{\mathrm{max}}=0.04$. The circled
  objects are the two absorbers discussed by
  \citet{2008MNRAS.391L..69S}, which are found to contain signatures
  of the 2175\ang bump, and the solid box indicates objects which are
  very likely to contain bump signatures based upon their LMC/SMC
  $D_{\mathrm{max}}$ statistic.}
\label{cap:natdustcomp2}
\end{figure}

\subsubsection{Results from the composite spectrum}\label{sec:comspec}
We created a composite spectrum (Fig \ref{cap:coaddstrongmg}) from
\compnum \, objects satisfying the condition in Equation
\ref{ebvcond}, by taking the mean\footnote{The mean (rather than the
  median) is now used to generate the composite spectrum to allow us
  to estimate the contribution of SMC-/LMC-/MW-type extinction to the
  composite.} flux value at each wavelength. Three further composites
were created from the fitted extinction curves (SMC, LMC, MW)
corresponding to each absorber. This allows for the characterisation
of the type of dust by investigating whether the stacked spectrum is
well fit by the SMC, LMC or MW curves. The resulting curves possess
$E(B-V)$ values of 0.12, 0.16, and 0.13 for the SMC, LMC and MW
respectively. As predicted, Fig \ref{cap:coaddstrongmg} shows that on
average our carefully-selected high $E(B-V)$ absorber subsample
exhibits an extinction curve very similar to that of the SMC.

We can quantify the significance of the SMC- versus LMC- and MW-type
reddening by examining the relative contributions of SMC-, LMC- and
MW-type dust to the composite spectrum in the region around the
2175\ang feature (Fig. \ref{cap:coaddstrongmg}-(\emph{bottom-right})).
We obtain an estimate for the ratio of
SMC-to-LMC-to-MW type spectra which contribute to our composite, by
coadding a number of SMC, LMC, and MW extinction curves and minimising
the residuals\footnote{The squared deviations between the underlying
  composite spectrum and the coadded extinction curves.} between the
coadd and the composite spectrum (in the range
1900--2500\,\AA). Formally, we found that a combination of \smcspec
\ SMC-, \lmcspec \ LMC- and \mwspec \ MW-curves provides the best fit
to the composite spectrum in the 2175\ang feature region. The
best-fitting combination is shown in
Fig. \ref{cap:coaddstrongmg}-(\emph{bottom-right}). The analysis shows
that of the absorbers which satisfy Equation \ref{ebvcond}, their
extinction curves are well reproduced by a mix of SMC and LMC
extinction curves in the ratio 65:35 with an uncertainty of
$\simeq$$\pm$10 in the ratio.

\begin{figure*}
\includegraphics[width=0.8\textwidth]{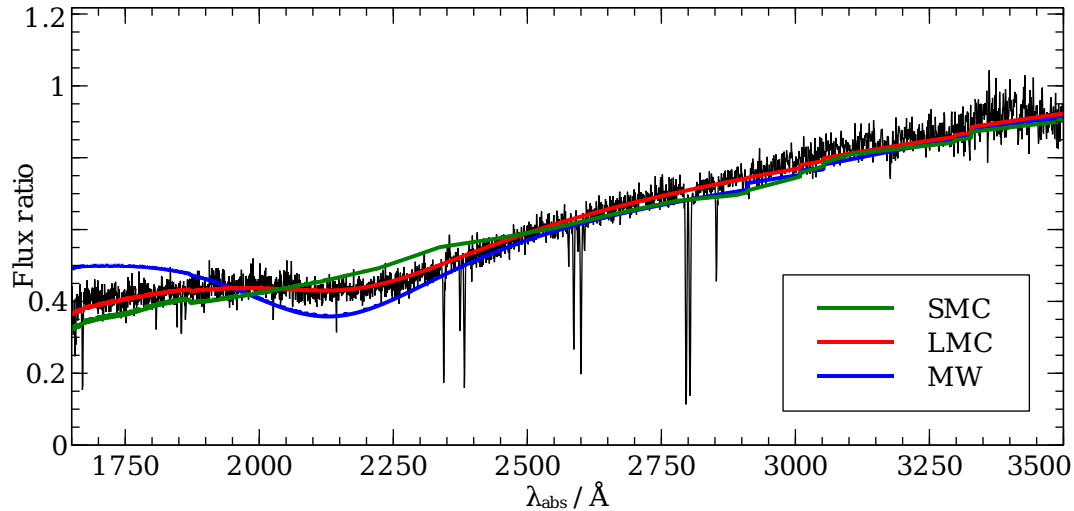}
\caption{The flux ratio against wavelength for a composite mean
  spectrum created from 18 objects satisfying Equation
  \ref{ebvcond}. The absorbers have been chosen (using the
  $D_{\mathrm{max}}$ statistic) as those most likely to exhibit
  LMC-type dust. The underlying composite spectrum is best fit by the
  LMC composite extinction curve, with mean $E(B-V)_{\mathrm{SMC}}=0.16$, $E(B-V)_{\mathrm{LMC}}=0.21$ and 
$E(B-V)_{\mathrm{MW}}=0.18$.}
\label{cap:lmccoadd}
\end{figure*}

The existence of absorbers with LMC-type dust can be demonstrated more
directly via consideration of Fig. \ref{cap:natdustcomp2}, which shows
the relationship between $D_{\mathrm{max}}^{\mathrm{SMC}}$ (x-axis)
and both $D_{\mathrm{max}}^{\mathrm{LMC}}$ and
$D_{\mathrm{max}}^{\mathrm{MW}}$ (y-axis) for the 211 absorber
systems. in general, $D_{\mathrm{max}}^{\mathrm{LMC}} <
D_{\mathrm{max}}^{\mathrm{MW}}$, which is to be expected if the
majority of absorbers contain SMC- or LMC-type dust. Coadding the flux
ratio spectra for the subset of 18 absorbers with
$D_{\mathrm{max}}^{\mathrm{SMC}}>0.02$ and
$D_{\mathrm{max}}^{\mathrm{LMC}}<0.02$ produces a composite flux ratio
(Fig. \ref{cap:lmccoadd}) the shape of which is closely reproduced by
the LMC extinction curve. The mean $E(B-V)$'s for the SMC, LMC, and MW
type extinction curves in this composite sample are 0.16, 0.21, and
0.18 respectively. It can be argued that the result must be true by
construction, with individual objects chosen to possess LMC-like
extinction curves. However, the procedure is no different from the
identification of individual systems \citep[e.g.]{2008MNRAS.391L..69S}
from among large samples of quasar spectra
\citep[e.g.]{2008MNRAS.391L..69S}.  More convincingly, the 18 systems
represent a significant minority of the base sample of
211. Furthermore, only a tiny fraction of absorption systems were
eliminated from the sample via a $D_{\mathrm{max}}$ threshold
(Fig.~\ref{cap:ks}) and it is very unlikely that unrelated intrinsic
quasar-to-quasar SED variations are combining to reproduce a 2175\ang
feature and overall shape that reproduces the form of the
LMC-extinction curve to the accuracy shown in Fig.~\ref{cap:lmccoadd}.
The objects contributing to the composite are indicated in
Fig. \ref{cap:ebvz}. With only half of these objects present in the
restricted sample used to investigate redshift evolution (Section
\ref{sec:redevol}), it is not possible to say anything useful about
the strength of the 2175\ang bump with redshift.

\citet{2006MNRAS.367..945Y} perfomed a statistical analysis of 809 \mg \ 
absorption systems from SDSS DR1, and searched for the 2175\ang
feature in the dustiest 111 systems with an observed frame
colour-excess\footnote{This excess corresponds to the difference
between the actual colours of a quasar and the median colours of a
quasar at that redshift.} $\Delta \left(g-i\right)>0.2$. They find an
average $E(B-V)_{\mathrm{SMC}}=0.0805$ and no evidence of a MW dust signature
in their composite spectrum. They also provide an estimate of the
fraction of lines of sight that could have MW-type extinction, and
provide an upper limit of 30 per cent.

Although we have reached a similar conclusion regarding the nature of
dust, our sample of \mg \ absorbers is an order of magnitude larger,
and we have created a composite spectrum with a much larger average
$E(B-V)_{\mathrm{SMC}}$. The sample used has also been carefully
constructed to ensure that only objects with accurate estimates of
$E(B-V)$ (i.e. those with $D_\mathrm{max}^{\mathrm{smc}}<0.04$) are
allowed to contribute to the composite. We provide a tighter
constraint on the fraction of lines of sight with (the much less
extreme) LMC-type extinction, as 35$\pm10$ per cent, compared to
\citeauthor{2006MNRAS.367..945Y}'s upper limit of 30 per cent with
MW-type extinction.

A recent study by \citet{2011ApJ...732..110J} produced a catalogue of
39 candidate strong \mg \ absorbers, with EW$>1$\ang and
$1.0<z<1.86$, showing evidence for the presence of a weak 2175\ang
feature.  We find 19 of the \citet{2011ApJ...732..110J} objects in our
sample. All possess moderate $E(B-V)_{\mathrm{SMC}}\sim0.1$, and 15/19
satisfy the condition
$D_\mathrm{max}^{\mathrm{lmc}}<D_\mathrm{max}^{\mathrm{smc}}$.
\citet{2011ApJ...732..110J}'s highlighting of these absorbers adds
further weight to the conclusion that a significant fraction of
absorbers with intermediate $E(B-V)$s show evidence for the presence
of a weak 2175\ang feature.

\subsubsection{The dustiest absorbers and the shape of the extinction curve}\label{sec:dustiest}

Very few absorbers at cosmological distances with unambiguous 2175\ang
features have been discovered. As discussed above,
\citet{2008MNRAS.391L..69S} find two \mg \ absorbers at $z$$\sim$1.3,
which are best-fit by LMC-type extinction, with
$E(B-V)_{\mathrm{LMC}}$'s of 0.27 and 0.36 respectively. Our analysis
also finds these absorbers to be best-fit by LMC-type extinction, with
corresponding $E(B-V)_{\mathrm{LMC}}$'s of 0.30 and 0.36, consistent
with their result. The two objects are present in our composite
subsample and are indicated in Figs. \ref{cap:ebvz} and
\ref{cap:natdustcomp2}. More recently, \citet{2010ApJ...724.1325J}
find two \mg \ absorbers at $z$$\sim$1.4, which may show detectable
2175\ang features. Our analysis finds the spectrum of
SDSS~J012147.73+002718.7 to be well fit by an LMC extinction curve
with $E(B-V)_{\mathrm{LMC}}=0.17$, whereas the other spectrum
(SDSS~J145907.19+002401.2, henceforth J1459+0024), is best fit by the
SMC curve with
$E(B-V)_{\mathrm{SMC}}=0.31$. \citeauthor{2010ApJ...724.1325J} find
that J1459+0024 is not particularly well fit by any SMC, LMC or MW
models because the extinction bump is much broader than that of the
average LMC extinction curve. A detailed discussion of the shape of
the extinction curve for J1459+0024, and the extinction curve of other
high $E(B-V)_{\mathrm{SMC}}$ objects is presented below.

Focussing on absorbers with comparable, or greater, estimates of
$E(B-V)$, assuming an SMC-type extinction curve, there are 25 systems
with $E(B-V)_{\mathrm{SMC}}$$\ge$0.30. Unsurprisingly, the majority of
absorbers have relatively low redshifts, $z_{\rm abs}$$\la$1, although
a small number extend out to nearly $z_{\rm abs}$=1.5. Two of the
systems are present in the sample of absorbers with detectable \caii \, 
absorption discovered by \citet{2005MNRAS.361L..30W} and
\citet{2006MNRAS.367..211W}.  The most extreme system, with
$E(B-V)_{\mathrm{SMC}}$=0.60, has been the subject of an earlier study.
\citet{2010ApJ...708..742Z}'s analysis of SDSS~J100713.68+285348.4
highlights the strength of the 2175\ang feature in the strong \mg \
absorber, at $z_{\rm abs}$=0.884, which also shows extremely strong
\caii \, absorption, although they do not note the connection to the
\caii \, absorber population. Employing constraints on the shape of the
extinction curve extending to the rest-frame near-infrared, provided
by $JHK$ photometry from 2MASS \citep{2006AJ....131.1163S},
\citet{2010ApJ...708..742Z} present an extinction curve using the
parametrization of \citet{2007ApJ...663..320F}. The form of the curve
includes a very strong 2175\ang feature, superposed on a rather
greyer background compared to any of the SMC, LMC or MW extinction
curves.  \citet{2010ApJ...708..742Z}'s fit to the SDSS spectrum and
the broadband photometry leads to a smaller value of $E(B-V)$=0.28 but
with a relatively large value of $R_{V}$=3.87.

Using our full sample of 25 objects with
$E(B-V)_{\mathrm{SMC}}$$\ge$0.30, it is possible to gain further
insight into the shape of the extinction curve associated with the
dustiest intervening absorbers.  Table \ref{tab:25objs} provides a
summary of information for the 25 systems and the background quasars,
including $YJHK$ photometry from the UKIDSS survey
\citep{2007MNRAS.379.1599L} where available. The population is not
significantly influenced by the effects of extinction from two or more
absorbers along the line of sight to a quasar.  An extensive search
for strong absorbers, including the \fetwo \ line complexes at
$\lambda$2344.2 and $\lambda$2586.7, covering the redshift interval
0.36$<$$z$$<$2.88, shows six of the quasars possess two absorbers with
$W_0^{\lambda2796}$$>$1.0\,\AA, whereas three are expected by chance.

A striking feature of the absorption line spectra is the almost
ubiquitous presence of \caii \, $\lambda\lambda$3934.8,3969.6
absorption. Fourteen of the 23 absorbers with $z_{\rm abs}$$<$1.31,
where \caii \, lies within the SDSS spectra, individually show
evidence for \caii \, absorption and a median composite of the 23
absorbers shows a rest-frame EW of \caii \,
$\lambda$3934.8=$0.38\pm0.05$\,\AA. The absorption systems are thus
closely related to the (observationally) rare, dusty \caii \, absorber
population identified by \citet{2005MNRAS.361L..30W}, the properties
of which are discussed by \citet{2006MNRAS.367..211W,
  2007MNRAS.374..292W}.  The presence of strong absorption due to a
species that is so strongly depleted onto dust as \caii \, is most
naturally explained by the presence of very large gas columns
\citep{2006MNRAS.367..211W} and our interpretation is that these
absorbers will possess hydrogen columns as large as DLAs.

Using a model unreddened quasar spectrum \citep{2008MNRAS.386.1605M}
we essentially confirm the results of \citet{2010ApJ...708..742Z} for
the shape of the extinction curve for the absorber in SDSS
J100713.68+285348.4. It is possible to undertake an identical analysis
for the six additional absorbers that possess $YJHK$
photometry\footnote{The faintness of the quasars, combined with the
  bright limit to the 2MASS photometry, means that choosing to analyse
  just those quasars that are detected in 2MASS (at low S/N) results
  in a significant bias, making the detected subset artificially red
  in $i-K$ and we thus confine our attention to quasars with UKIDSS
  photometry.}. In all cases, adopting our $E(B-V)$ values (using any
of the SMC, LMC or MW curves) from the SDSS spectra produces
predictions that are grossly inconsistent with the SDSS ($i$-band) to
$YJHK$ colours. Specifically, compared to any of the
SMC/LMC/MW-extinction curves, the shape of the extinction curve probed
by the SDSS spectra of the absorbers at redshifts $z_{\rm
  abs}$$\simeq$0.75 (rest-frame wavelengths $\simeq$2100-5000\,\AA) is
much steeper than the curve at rest-frame wavelengths
$\simeq$4500-14\,000\ang probed by the broadband $i$ to $K$
colours. In other words, the extinction curve is significantly greyer
at $>$5000\,\AA, with local extinction curves predicting a median
$i-K$ colour 0.72\,mag redder than that observed for the six
absorbers.

While the five-parameter parametrization of
\citet{2007ApJ...663..320F} would provide accurate fits to the SDSS
spectra and the $i$ through $K$ photometry of the six absorbers, we chose
to adopt the established one-parameter family of extinction curves
presented by \citet{1999PASP..111...63F}, in which the shape of the
untraviolet to near-infrared extinction curves depends on the value of
$R_V$ (see fig. 7 of \citep{1999PASP..111...63F}).

We find satisfactory fits to both the flux ratio spectra derived from
the SDSS quasar spectra and the $iYJHK$ broadband photometry of the
six objects using a \citet{1999PASP..111...63F} curve with $R_V$=2.1
(see Appendix \ref{sec:appendix}). The $D_{\mathrm{max}}$-statistic
distribution is as good as that from any of the SMC, LMC, MW or Zhou
et al. extinction curves. The median $i-K$ colours (model - observed)
using the Fitzpatrick and Zhou curves are +0.09 mag and -0.04
respectively (cf. the +0.72 value using any of the SMC, LMC or MW
curves).  Individual $iYJHK$ magnitudes are reproduced to the level of
$\sigma \simeq0.12$\,mag using either the Fitzpatrick or Zhou curves
but with some differences reaching 0.3\,mag. The small number of
absorbers, combined with the effects of intrinsic quasar variability
(between the epochs of the SDSS and UKIDSS photometry) and object to
object SED variation, preclude the generation of more exact
constraints on the form of the extinction curve but the need for a
significantly greyer shape in the near-ultraviolet and optical is clear.

The inferred values of $E(B-V)$ for the six systems employing the
Fitzpatrick $R_V$=2.1 curve are given in Table \ref{tab:25objs}. We
note that the form of the Fitzpatrick $R_V$-dependent extinction
curves at low values of $R_V$ reproduces the shape of the extinction
curve in the near ultraviolet with a 2175\ang feature of strength
comparable to that in the MW, combined with a somewhat steeper
increase in extinction with decreasing wavelength, a key feature of
the behaviour found by \citet{2007ApJ...663..320F}.  From the data
available, it is not possible to conclude whether
\citet{2010ApJ...708..742Z}'s modeling, employing a super-strong
2175\ang feature, or the $R_V$-dependent form (with an extreme value
of $R_V$=2.1) of Fitzpatrick (1999), is superior but the simple
one-parameter-dependent form, with a 2175\ang feature of constant
strength, is attractive. In the Appendix we provide both a graphical
representation of the extinction curve and quantitative values of
$E(\lambda-V)/E(B-V)$ as a table.

\begin{figure*}
\includegraphics[width=0.8\textwidth]{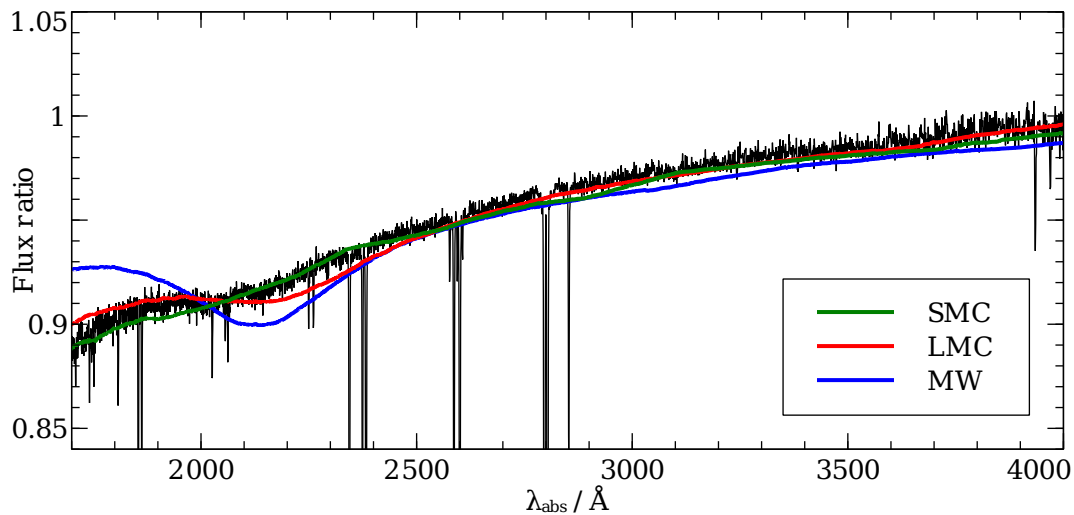}
\caption{The flux ratio against wavelength for a composite median
  spectrum created from 1\,792 absorbers with $0 \leq
  E(B-V)_{\mathrm{SMC}} \leq 0.05$ and $z_{\mathrm{abs}}>0.9$. The
  shape of the flux ratio spectrum for the low $E(B-V)$ composite is
  consistent with SMC-type extinction with median
  $E(B-V)_{\mathrm{SMC}}=0.018$, $E(B-V)_{\mathrm{LMC}}=0.022$ and $E(B-V)_{\mathrm{GAL}}=0.016$. The minimum flux ratio on the y-axis
  has deliberately been set to 0.84 and the x-axis extended to 4100\ang
  to better show the shape of the extinction curve at small deviations
  from unit flux-ratio. A composite based on the mean flux-ratio for
  the sample is noisier but extremely similar in shape.}
\label{cap:smccoadd}
\end{figure*}

The difference in the form of the extinction curve (from SMC/LMC/MW
curves) for absorbers with high $E(B-V)$, including the likely
presence of a strong 2175\ang feature, raises the question of whether
there is a systematic dependence of the extinction curve shape with
$E(B-V)$.  While the detection of the 2175\ang feature in individual
spectra with small $E(B-V)$ is not viable, the prevalence of an
SMC-like curve for absorbers with low extinctions ($E(B-V)$$<$0.05) is
unambiguous; any extinction curve with a 2175\ang feature as strong as
that in the MW can be ruled out with confidence
(Fig. \ref{cap:smccoadd}). For absorbers with intermediate
$E(B-V)$$\simeq$0.15 we find evidence for the presence of a weak
2175\ang feature in a significant fraction of systems (Section
\ref{sec:comspec}). The results for systems with larger
$E(B-V)_{\mathrm{SMC}}$ require confirmation using a much larger
sample of absorbers with near-infrared photometry but for the first
time there is an indication of a strong systematic change in the form
of the near-ultraviolet to near-infrared extinction curve as a
function of $E(B-V)$, with the shape of the curve in the dustiest
systems differing from any of the standard curves in the local
Universe.

\begin{table*}
\caption{A sample of 25 objects with $E(B-V)_{\mathrm{SMC}}>0.3$}
\begin{minipage}{190mm}
\label{tab:25objs}
\begin{tabular} { lcllccccccc } 
\hline
SDSS name & $z_\mathrm{qso}$ & $z_\mathrm{abs}$ & $W_{0}^{\lambda  2796}$ & $E(B-V)_{\mathrm{SMC}}$ & $E(B-V)_{\mathrm{Fitz}}$ & $i_{\mathrm AB}$ & $Y_{\mathrm Vega}$ & $J_{\mathrm Vega}$  & $H_{\mathrm Vega}$ & $K_{\mathrm Vega}$ \\
          &                 &                 & (\AA)                &                       &                        &                 &                 &                   &                  &                   \\
\hline 
J081524.62+555153.3 & 1.781 & 1.323(1.608) & 1.9(1.3) & 0.37 & 0.31 & 18.63 & --- & --- & --- & ---\\ 
J083257.64+333214.6 & 1.017 & 0.716 & 2.2 & 0.33 & 0.29 & 19.06 & --- & --- & --- & ---\\ 
J085244.74+343540.4$^a$ & 1.656 & 1.310(0.884) & 3.1(2.5) & 0.31 & 0.27 & 18.21 & --- & --- & --- & ---\\ 
J091753.89+000300.8 & 2.139 & 0.729 & 4.3 & 0.34 & 0.31 & 18.78 & 17.92 & 17.46 & 16.69 & 15.84\\ 
J092053.12+385002.7 & 1.276 & 0.472 & 1.8 & 0.30 & 0.28 & 17.96 & --- & --- & --- & ---\\ 
J092333.29+595747.8 & 0.839 & 0.662 & 2.6 & 0.31 & 0.28 & 19.02 & --- & --- & --- & ---\\ 
J093444.30+172453.5 & 1.899 & 0.553 & 2.8 & 0.33 & 0.32 & 19.00 & --- & --- & --- & ---\\ 
J093738.04+562838.9$^b$ & 1.804 & 0.978 & 4.9 & 0.30 & 0.23 & 18.50 & --- & --- & --- & ---\\ 
J100713.68+285348.4$^c$ & 1.050 & 0.884 & 3.5 & 0.60 & 0.50 & 18.25 & --- & --- & --- & ---\\ 
J113152.15+435318.3 & 2.139 & 1.098(1.377) & 2.4(1.1) & 0.34 & 0.28 & 18.82 & --- & --- & --- & ---\\ 
J113811.59+382119.5 & 0.899 & 0.759 & 1.9 & 0.44 & 0.38 & 18.41 & --- & --- & --- & ---\\ 
J120301.00+063441.5$^d$ & 2.180 & 0.862 & 5.6 & 0.42 & 0.37 & 18.44 & 17.28 & 16.98 & 16.41 & 15.51\\ 
J120913.61+433920.9 & 1.397 & 0.412 & 2.4 & 0.30 & 0.28 & 17.35 & --- & --- & --- & ---\\ 
J121547.11+293409.9 & 1.532 & 0.788 & 1.0 & 0.30 & 0.25 & 17.88 & --- & --- & --- & ---\\ 
J124946.59+120400.0 & 1.688 & 1.469 & 2.3 & 0.33 & 0.31 & 18.20 & 16.90 & 16.28 & 15.49 & 15.07\\ 
J131103.19+551354.3 & 0.927 & 0.600 & 1.2 & 0.39 & 0.37 & 18.87 & --- & --- & --- & ---\\ 
J133753.11+052746.2 & 1.743 & 0.575 & 1.4 & 0.31 & 0.30 & 17.99 & --- & --- & 16.05 & 15.72\\ 
J140807.05+374457.3 & 1.497 & 0.776 & 3.0 & 0.30 & 0.25 & 18.61 & --- & --- & --- & ---\\ 
J144611.68+484613.7 & 1.202 & 0.669 & 2.2 & 0.38 & 0.34 & 18.46 & --- & --- & --- & ---\\ 
J144642.92+012552.4 & 1.422 & 0.525(0.510) & 2.4(1.4) & 0.34 & 0.32 & 18.08 & 17.48 & 16.92 & 16.13 & 15.71\\ 
J145344.23+102557.5 & 1.770 & 0.757 & 2.6 & 0.38 & 0.34 & 19.02 & 18.09 & 17.66 & 17.10 & 16.40\\ 
J145907.19+002401.2$^e$ & 3.037 & 1.394 & 1.7 & 0.31 & 0.23 & 17.83 & --- & --- & --- & ---\\ 
J155435.08+484411.1 & 2.041 & 0.912(1.385) & 2.4(1.1) & 0.31 & 0.24 & 18.98 & --- & --- & --- & ---\\ 
J170220.06+591538.6 & 1.798 & 0.724 & 1.8 & 0.32 & 0.28 & 18.73 & --- & --- & --- & ---\\ 
J171123.03+311613.7 & 2.041 & 1.256(1.697) & 2.3(1.9) & 0.31 & 0.27 & 18.65 & --- & --- & --- & ---\\ 
\hline 
\end{tabular}

\medskip
Magnitudes are corrected for Galactic extinction using the
prescription of \citet{1998ApJ...500..525S}.  Photometric errors for
the SDSS $i$-band photometry are normally $<0.03$\,mag and for the
UKIDSS photometry $\simeq$0.05\,mag.  The absorber redshifts and EWs
in brackets correspond to those spectra which have two \mg \ systems with
EW$>$1\,\AA. Errors in the rest-frame EWs are typically
$\pm$0.1\,\AA. $E(B-V)_{\mathrm{Fitz}}$ is the result of the fit to
the extinction curve of \citet{1999PASP..111...63F} with $R_V$=2.1
(see Appendix~\ref{sec:appendix}).  Systems that have been the subject
of previous investigations are designated ($^a$)
\citet{2008MNRAS.391L..69S},($^b$) \cite{2005MNRAS.361L..30W}, ($^c$)
\citet{2010ApJ...708..742Z},($^d$) \cite{2006MNRAS.367..211W} and
($^e$) \citet{2010ApJ...724.1325J}.

\end{minipage}
\end{table*}

\subsection{Implications for M\lowercase{g}\,{\sevensize\bf II} 
absorber statistics}\label{mgstatssec}
\begin{figure*}
\includegraphics[width=0.4\textwidth]{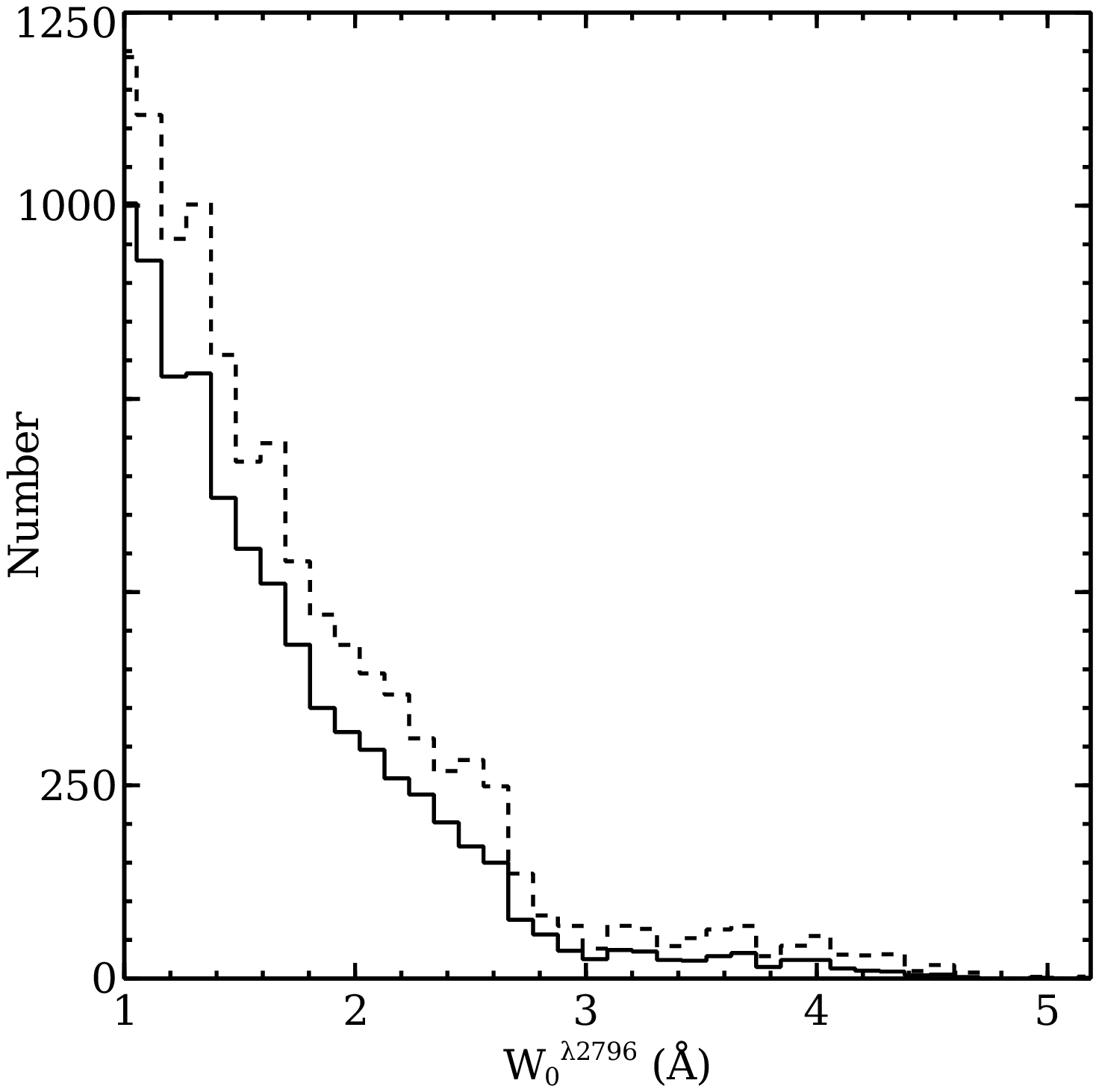}
\hspace{5mm}
\includegraphics[width=0.4\textwidth]{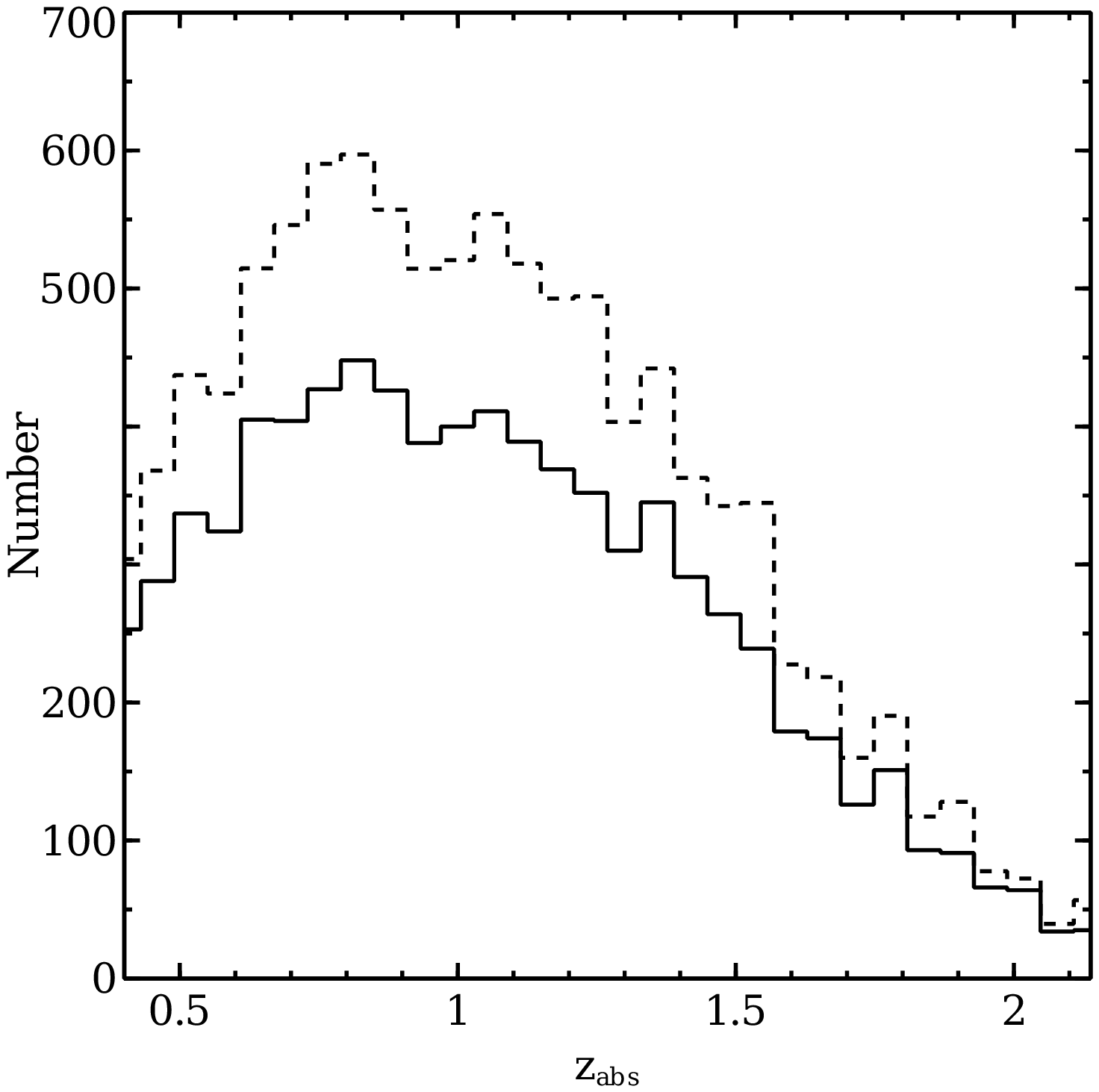}
\caption{Left-hand panel: The EW distribution of the absorber sample
  (solid-line), and corresponding dust-corrected EW distribution
  (dashed-line), which has been calculated using
  equation~\ref{eq:weightebv_old}. Right-hand panel: The redshift
  distribution of the absorber base sample (solid-line), and the
  corresponding dust-corrected distribution (dashed-line).}
\label{cap:cordists}
\end{figure*}

The quantitative dust-dependent bias, formulated in terms of the
absorber $E(B-V)$, derived in this paper has significant implications
for the study of the \mg \ absorber
population. Fig. \ref{cap:cordists} shows the distribution of observed
EW and redshift for our absorber sample. Also shown are the
bias-corrected distributions, calculated using the object by object
bias-corrections (equation \ref{eq:weightebv_old}). A significant
fraction of absorbers are missing from the sample, \mgmissperc \ $\pm$
\mgmisspercerr \ per cent for absorbers with EW$>$1.0\ang in the
redshift interval $0.4<z<2.2$. The missing fraction rises to
\mgmissperchigh \ $\pm$ \mgmissperchigherr \ per cent for absorbers
with EW$>$2.0\,\AA. The formal errors on the incompleteness fraction
are at the level of 1 per cent and the error is dominated by the
systematic uncertainty associated with the dc offset (Section
\ref{sec:acccors}). The systematic uncertainty becomes less significant for
the EW$>2.0$\ang sample.

\citet{2005ApJ...628..637N} find the the EW distribution of a sample
of 1\,300 absorbers (with redshifts $0.366\leq z \leq2.269$) from SDSS
DR4 is well described by an exponential distribution. They find that
although moderately strong lines ($0.4\,\mathrm{\AA}<W_{0}^{\lambda
  2796}<2.0\,\mathrm{\AA}$) show no evidence for redshift evolution,
the absorbers with larger EWs show an increase in number with redshift
that is more pronounced for absorbers of higher EW.  Nestor et
al. interpret this redshift dependence as an evolution in the
kinematic properties of absorbers over a range of intermediate
redshifts. \citet{2009ApJ...698..819L} find a similar result which
suggests stronger evolution in the redshift number density of strong
\mg \ absorption systems relative to lower EW samples, albeit over a
smaller redshift range $0.4 \leq z \leq 0.8$.

Using our parametrized dust correction (equation \ref{eq:fullebvew}),
we predict that the \emph{observed} relative fraction of high EW
systems will decrease by \mgfrac \ per cent over the redshift interval
$0.4 < z < 2.0$. Our results thus predict that the true increase in
the proportion of high EW systems with increasing redshift is
significantly greater than found by \citet{2005ApJ...628..637N} from
the \emph{observed} population of absorbers.

\subsection{The GRB/Quasar absorber number density sight line discrepancy}

We have discussed the effects of dusty high EW absorbers on the
completeness of optical magnitude-limited quasar samples over a range
of absorber redshifts. The highest redshift absorbers have been
observed using GRB optical afterglows as background sources
\citep{2009GCN..9215....1O}. Although GRBs would individually
experience the same obscuration effects, GRBs are found to have a huge
range in intrinsic brightness and spectra are obtained at variable
time intervals following the GRB's peak brightness. The consequent very
large dispersion in apparent GRB brightness, rather than the
presence of modest amounts of dust in any intervening absorbers,
dictates whether an object is observed and absorption systems
identified. Therefore, one would expect a larger number of moderate
$E(B-V)$ absorption systems to be present in GRB spectra than quasar
spectra.

\citet{2006ApJ...648L..93P} identified 14 strong intervening \mg
\ systems (at a mean redshift of $\langle z \rangle=1.1$) along 14 GRB
sight lines, an incidence roughly four times higher than along sight
lines to quasars. The result is not expected if both GRBs and quasars
sample random lines of sight. Since the intervening absorption systems
are thought to be independent of the background source, the observed
discrepancy has led to a call to review the fundamental assumptions
that underpin extragalactic absorption line research. The discrepancy
was confirmed by \citet{2007ApJ...669..741S}, but the amplitude has
since been reduced to a factor of $2.1 \pm 0.6$ using a larger sample
of 22 absorbers \citep{2009A&A...503..771V}.

A series of explanations have been suggested to explain the observed
GRB/quasar discrepancy. \citet{2007ApJ...659..218P} claim that strong
\mg \ absorbing gas may be intrinsic to the GRB circumburst
environment or originate from supernova remnants lying in the same
star-forming region. It is also claimed \citep{2006ApJ...648L..93P,
  2007ApJ...659..218P, 2009ApJ...706.1309T} that source amplification
due to strong gravitational lensing may bias the GRB spectral samples
toward targets that contain more intervening absorbers. Another
mechanism has been proposed \citep{2007Ap&SS.312..325F}, which claims
that the discrepancy is due to the different beam sizes of GRBs and
quasars, but this has subsequently been ruled out by observational
analysis \citep{2007MNRAS.381L..99P}. Finally, a number of authors
\citep{2006ApJ...648L..93P, 2007ApJ...659..218P} have suggested that
dust associated with strong \mg \ systems results in a reduction in
the observed number of absorbers seen along sight lines to quasars in
flux-limited catalogues. In general, most authors agree that the
differences between \mg \ toward quasar and GRB sight lines cannot be
due to a single effect, but dusty absorbers have not been thought to
be a significant factor.

In light of our findings on the existence of a significant population
of dusty \mg \ absorbers, we can provide a more definitive
determination of the effect of dust on the GRB/quasar sight line
discrepancy. Using a sample of GRB sight lines with intervening
absorbers we can assess how this sample would suffer from
incompleteness if the sight lines were instead illuminated by
quasars. We assume that the GRB sample is unbiased and probes all
intervening \mg \ absorbers up to a moderate absorber $E(B-V)$ of 0.2
mag. Given an EW and redshift distribution of \mg \ absorbers in GRBs
similar to our quasar sample (Fig. \ref{cap:cordists}), we would
expect the GRB sight lines to show the missing \mgmissperc \ per cent
of absorbers (Section \ref{mgstatssec}). However, in general GRB
spectra contain absorbers with different redshift and EW distribution
to the absorbers in quasar spectra. The observed discrepancy in number
density due to dust will be sensitive to such differences.

\begin{figure*}
\includegraphics[width=0.4\textwidth]{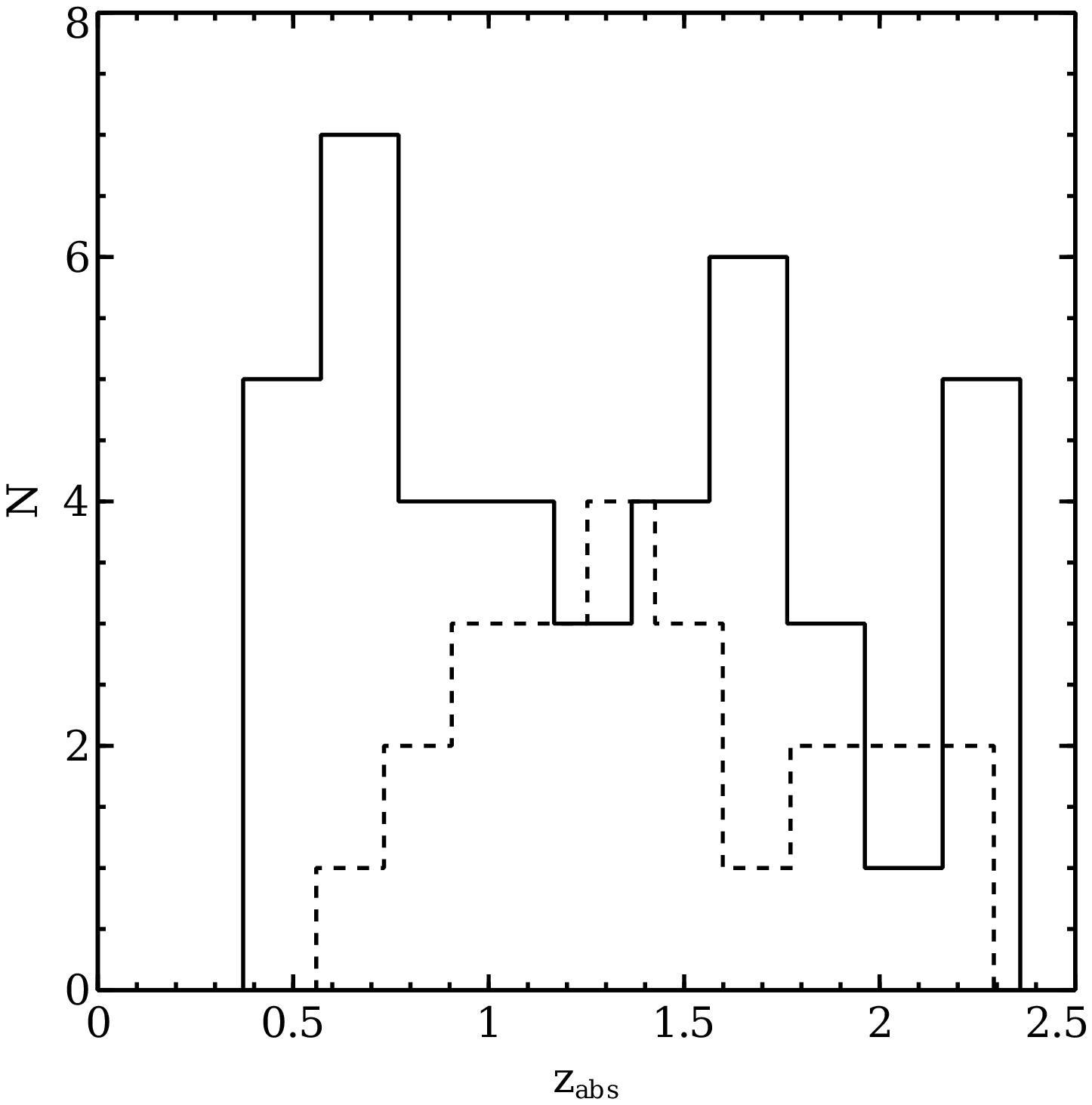}
\hspace{5mm}
\includegraphics[width=0.4\textwidth]{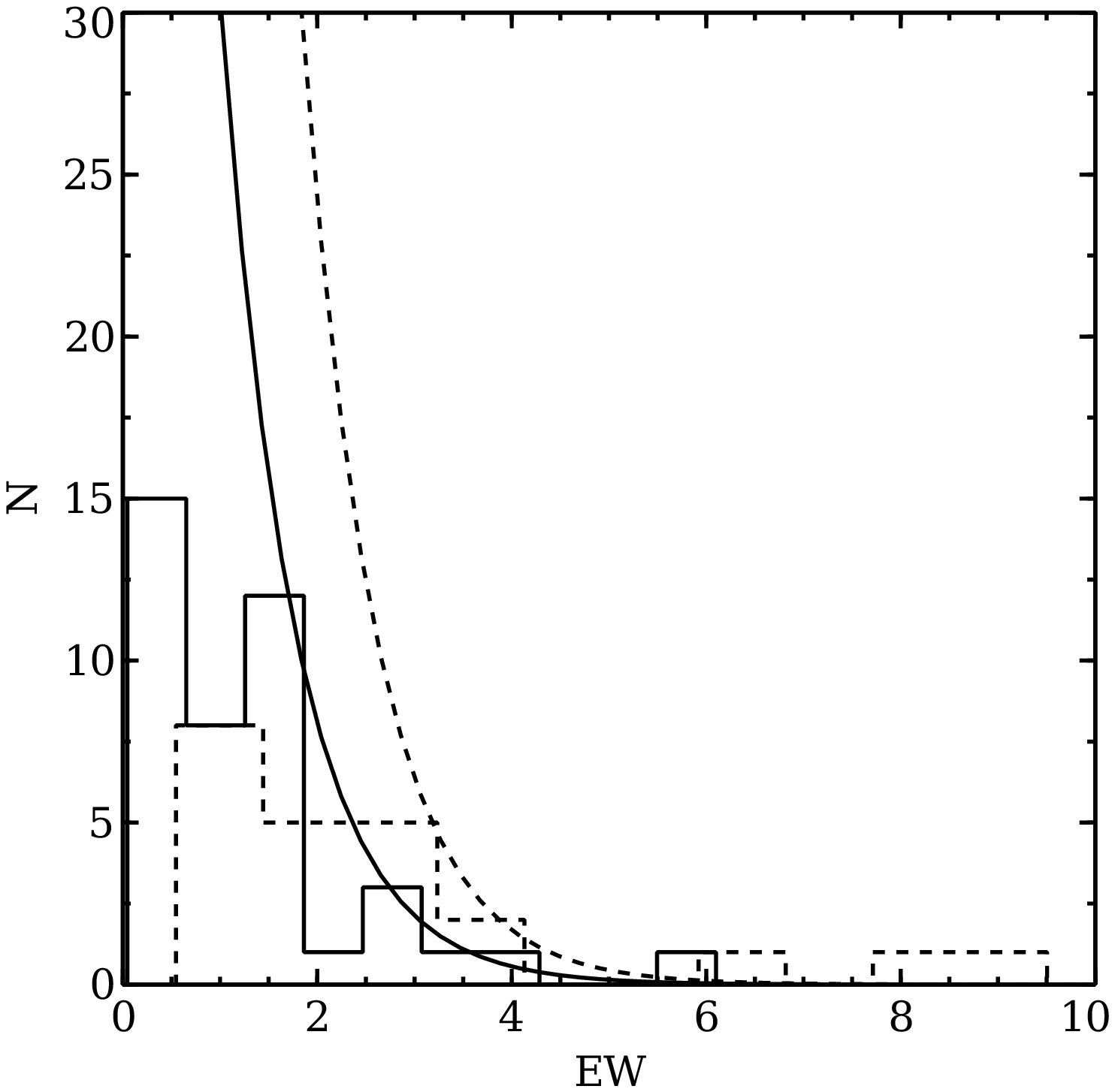}
\caption{Left-hand panel: The redshift distribution of the
  \citeauthor{2009A&A...503..771V} GRB absorber sample (solid-line),
  and the \citeauthor{2009ApJS..185..526F} sample (dashed-line). The
  absorbers exhibit similar redshift coverage, although the
  \citeauthor{2009ApJS..185..526F} sample shows a slightly larger
  fraction of absorbers at higher redshift than the
  \citeauthor{2009A&A...503..771V} sample. Right-hand panel: The EW
  distribution of the \citet{2009A&A...503..771V} GRB absorber sample
  (solid-line), and the \citet{2009ApJS..185..526F} sample
  (dashed-line). Overlayed are the results of an exponential fit
  (Equation \ref{nestorexp} with $W^{*}=0.746$) to the two EW
  distributions. The vertical scaling is calculated using the
  redshift path of the two samples. Both samples are expected to be
  highly incomplete at low EW. The extension of the absorber
  distribution to $W_{0}^{\lambda2796}$$\simeq$10\ang in the 
  \citeauthor{2009ApJS..185..526F} sample is clear.}
\label{cap:grbzewdists}
\end{figure*}

Two GRB absorber samples in the literature are the
\citet{2009A&A...503..771V} sample and the \citet{2009ApJS..185..526F}
sample. The \citeauthor{2009A&A...503..771V} sample contains 22 strong
systems (EW$>1$\,\AA) identified using high-resolution UVES sample and
a series of high- and low-resolution GRB afterglow spectra from the
literature. The \citeauthor{2009ApJS..185..526F} sample contains 15
strong systems present in the follow-up spectra of 77 optical
afterglows of \emph{Swift} detected GRBs. The redshift distributions
of the two samples are shown in
Fig. \ref{cap:grbzewdists}-(\emph{left}). The redshift distributions
for the two GRB samples both exhibit a larger fraction of absorbers at
high redshift compared with the quasar absorber redshift distribution.
Fig. \ref{cap:grbzewdists}-(\emph{right}) shows the degree to which
both samples suffer from EW incompleteness by comparing the EW
distributions with an exponential \citep{2005ApJ...628..637N} of the
form

\begin{equation}
N\left ( W_{0}^{\lambda 2796} \right ) \propto e^{-W_{0}^{\lambda 2796}/W^{*}}
\label{nestorexp}
\end{equation}

\noindent where $W^{*}$ is the exponential scale factor. A fit to the
corrected quasar absorber EW distribution (Fig. \ref{cap:cordists})
gives a value of $W^{*}$=0.746$\pm$0.008. Scaling the corrected quasar
EW distribution using the redshift paths appropriate to the two
GRB-derived samples results in the predicted number versus EW distributions
for the \citeauthor{2009A&A...503..771V} and
\citeauthor{2009ApJS..185..526F} GRB samples shown in
Fig. \ref{cap:grbzewdists}-(\emph{right}). Both samples are expected
to be incomplete at low EW but the scaled distribution functions
accurately describe the \citeauthor{2009A&A...503..771V} sample with
EW$\ga$2.0\,\AA, and the \citeauthor{2009ApJS..185..526F} sample is
consistent over the interval $\simeq$2.0--5.0\,\AA. Our flux-limited
quasar-derived \mg \ absorber distribution has no information
regarding high-EW systems with $W_{0}^{\lambda2796}$$>$5.0\ang (where
the associated large $E(B-V)$ essentially removes nearly all such
systems from the SDSS quasar sample).  The extension of the
distribution of \mg \ absorbers beyond $W_{0}^{\lambda2796}$=5.0 out
to $\simeq$10\ang is clear and strongly suggests that the absorber
dust-induced differences between quasar-derived and GRB-derived
samples will be even larger than calculated using our
$N(W_{0}^{\lambda2796})$ parametrization.

We can calculate the fraction of absorbers that would be missed in
corresponding flux-limited quasar sample by estimating an $E(B-V)$
according to equation \ref{eq:fullebvew} and assigning a weighting
according to the dust obscuration bias/signal-to-noise ratio effects 
(equation \ref{eq:weightebv_old}). The predicted discrepancy is
obtained by calculating the fraction of absorbers that are missed from
a quasar-derived sample with an $W_{0}^{\lambda2796}$ distribution corresponding
to Equation \ref{nestorexp} (with $W^{*}=0.746$) and redshift
distributions of the \citeauthor{2009A&A...503..771V} and
\citet{2009ApJS..185..526F} absorbers\footnote{Note that the procedure
  adopted is inherently conservative because the parametrized
  dependence of the $E(B-V)$ values on $W_{0}^{\lambda2796}$ reduces
  the impact of individual dusty absorbers.}.

\begin{figure}
\includegraphics[width=\columnwidth]{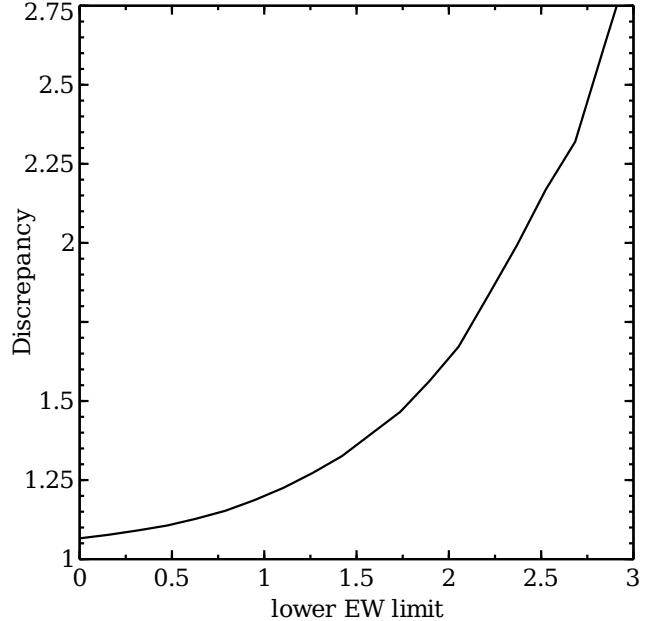}
\caption{The predicted discrepancy between GRB and flux-limited quasar
  sample sight lines due to dust as a function of lower
  $W_{0}^{\lambda2796}$ limit, for a GRB sample with EWs distributed
  according to Equation \ref{nestorexp} (with $W^{*}=0.746$), and
  redshift distribution consistent with that of the
  \citet{2009A&A...503..771V} and \citet{2009ApJS..185..526F}
  samples.}
\label{cap:discp}
\end{figure}

The strong dependence of the dust content on the EW of the \mg
\ absorbers means that the degree to which the incidence of absorbers
is affected is a very strong function of the low EW limit of the
samples (where the majority of absorbers reside). Fig. \ref{cap:discp}
shows the predicted discrepancies for a sample with the same absorber
redshift distributions as Vergani et al. (2009) and Fynbo et
al. (2009) as a function of the lower EW limit for the absorber
sample. The increasing impact of dust, as samples with increasingly
large lower EW limits are considered, illustrates the importance of
incorporating quantitative estimates when inter-comparing results from
different observed samples. The discrepancy reaches a full factor of
two for $W_{0}^{\lambda2796}$$>$2.4\,\AA.

Our results are consistent with the investigation of
\citeauthor{2009ApJ...706.1309T} who find no significant increase in
number density of low-EW absorbers, EW$<$1\,\AA, towards GRBs. For GRB
samples where high-EW absorbers (EW$>$2\,\AA) represent a significant
fraction of the absorber sample, our predicted enhancement in \mg
\ number density is a factor two compared to absorber samples seen
towards SDSS-selected quasars. The evidence from the
\citeauthor{2009ApJS..185..526F} sample for the existence of a
significant number of absorbers with $W_{0}^{\lambda2796}$s as high as
10\,\AA \ suggests that the true discrepancy will be significantly
greater than shown in Fig. \ref{cap:discp} for samples with
minimum detectable $W_{0}^{\lambda2796}$ above 1\,\AA.  The amplitude
of the enhancement is higher than has been reported in the literature
hitherto and dusty absorbers could be responsible for a significant
proportion of the absorber number density enhancement seen towards
GRBs.

A careful study of the incidence of \mg \ absorbers towards a third
class of objects, the BL Lacs (or Blazars), whose selection is closely
akin to GRBs in not being flux-limited in the optical, has been
conducted by \citet{2011A&A...525A..51B}. They find an excess of \mg \
absorbers along Blazar sight lines compared with quasar sightlines of
$\simeq$2.0 with no apparent difference between low-EW absorbers
(EW$<$1\,\AA), and high-EW absorbers (EW$>$1\,\AA). However, the absorber
sample is numerically small and while strong differences between low-
and high-EW absorbers are not present, less significant differential
effects as a function of absorber EW are not ruled out with any confidence.  

Our estimate of the true number density of strong (EW$>$1\,\AA) \mg
\ absorbers along sight lines to SDSS quasars is some 15 per cent
higher than the value used by \citet{2011A&A...525A..51B}, reducing
the observed overdensity somewhat.  The median redshift of the Blazar
absorber sample is low, even lower than for the SDSS quasars sample
presented here ($\bar{z}$=0.84 compared to $\bar{z}$=1.1), and the effect
of dusty absorbers on the statistics is correspondingly much lower
than for the GRB absorber samples discussed above. Our model predicts
an increased overdensity for the \citet{2011A&A...525A..51B} sample of
only 15 per cent.  We also note that the contribution to the
overdensity signal may be redshift dependent (fig. 2 of Bergeron et
al.) with the modest path-density at $z$$>$1.0 producing much of the
effect.  The \citet{2011A&A...525A..51B} study does not offer direct
support in favour of the predictions from our dusty absorber model but
nor does it rule out such a contribution. However, assuming that the
factor of two overdensity of absorbers at low redshifts for Blazar
samples is confirmed, then more contributors, in addition to the
effects of dusty absorbers, are required to explain the observed
absorber overdensity towards GRBs.

\section{Discussion and conclusions}\label{disc}
With the aid of the enormous statistical power of the SDSS, we have
been able to provide an improved characterisation of the dust content of
\mg \ absorption systems and its effects on the properties of absorber
samples obtained using magnitude-limited quasar surveys. The principal 
conclusions of our investigations are:

\begin{enumerate}
\item We find that dust content ($E(B-V)$) in a large sample of
$\simeq$8300 strong \mg \ absorbers increases much more strongly as a
function of absorber rest-frame equivalent width (EW) (see equation
\ref{eq:fullebvew}) than previously reported.
\item The effects of dust on the {\it observed} \mg \ absorber
population results in significant changes to the statistics of the
intrinsic absorber population. We find \mgmissperc \ $\pm$
\mgmisspercerr \ per cent of absorbers are missing from the sample for
EW $>1.0$\,\ang and \mgmissperchigh \ $\pm$ \mgmissperchigherr \ per
cent for EW $>2.0$\,\AA. For absorbers with EW $>3.0$\,\AA) the
fraction rises to more than a factor of two.
\item We find no significant detection of an evolution in absorber
  dust content over the redshift interval 0.4$\lid$$z$$\lid$2.2 but
  the constraint is relatively weak, with a 1$\sigma$ uncertainty of
  0.3 in the power-law index $\alpha$ ($E(B-V)$$\propto$($1+z$)$^\alpha$)
\item The form of the extinction curve for absorbers is found to be of
  SMC-type at low values of $E(B-V)$. For absorbers with modest
  $E(B-V)$$\la$0.2 at redshifts $z_{\mathrm{abs}}>1$, where the
  2175\ang is visible in the SDSS spectra, approximately a third of
  the population shows evidence for the presence of dust with an
  LMC-like extinction curve.
\item For absorbers with the highest $E(B-V)$$\ga$0.3 there is
  evidence for an extinction curve that differs from the SMC, LMC and
  MW extinction curves. The high $E(B-V)$ systems possess a 2175\ang
  feature similar in strength to that seen in the MW but the overall
  shape of the extinction curve in the near-ultraviolet through to the
  near-infrared, is greyer than any of the SMC, LMC or MW extinction
  curves.
\item The impact of dust on the observed discrepancy in the \mg
  \ absorber sight line densities between GRBs and quasars depends on
  both the absorber redshift distribution and the minimum
  equivalent width limit of the samples obtained towards GRBs. Samples
  with high lower EW-limits for the detection of \mg \ absorbers and
  significant redshift path for absorbers at high redshifts are most
  affected. Published GRB samples are predicted to possess absorber
  densities a full factor of two greater than for quasars from
  flux-limited samples due to the effects of dust extinction.
\end{enumerate}

Higher EW \mg \ systems are found to exhibit a higher star formation
rate \citep{2009arXiv0912.3263M} and the dependence of $E(B-V)$ on EW
can be explained in terms of the star formation rate of the absorbing
galaxy.  Supernovae associated with star formation impart mechanical
energy to the cold \mg \ absorbing gas, which leads to velocity
spreading of the absorption feature and a consequent increase in the
observed absorber EW. An increased amount of star formation also leads
to increased dust production. The presence of a
larger amount of dust in absorbers associated with the highest
star formation rate has potential consequences for determining the 
redshift-evolution of absorbers with the highest EWs although the
impact on the overall star formation rate density of the Universe
\citep{2004ApJ...615..209H} is unlikely to be more than a perturbation.

Explaining the observed differences between the \mg \ absorber density
towards quasars in flux-limited samples, GRBs and BL Lacs requires 
considerably larger absorber samples to map out the dependence of the
differences on absorber EW and redshift. While the explanation will
almost certainly involve more than one effect, the results presented here
show that obscuration due to dust associated with the absorbers is 
almost certainly a significant factor.

To date, the limited statistics available has meant that constraints
on the dust content and form of the extinction curves associated with
metal absorbers have been weak. The availability of metal absorber
samples with $\sim$10\,000 systems and the detection of a small number
of systems with $E(B-V)$$\ga$0.3 is changing the situation, as shown
in this paper.  At low values, $E(B-V)$$\la$0.05, the extinction
curve appears to be essentially featureless in the near-ultraviolet
with an SMC-like form. A new result from our study is the detection of
a weak 2175\ang feature in a significant fraction of \mg \ absorbers
with $E(B-V)$$\simeq$0.1-0.2, consistent with an LMC-like extinction
curve. Individual systems with higher $E(B-V)$ values of $\ga$0.3 and
an apparent strong 2175\ang feature have been identified in recent
years. Our results for the \mg \ absorbers with the highest $E(B-V)$
values, particularly the small subset with near-infrared photometry
available, indicate that there may be a systematic dependence in the
form of the extinction curve as a function of $E(B-V)$ and, more
fundamentally, the column density of the absorbers. Obtaining
near-infrared photometry, to combine with SDSS spectra, for a larger
sample of such objects is necessary to establish the general nature
of any such trend. Confirmation of a systematic
change in the form of the extinction curve as advocated here, from
featureless and SMC-like at low $E(B-V)$, through to greyer
overall than any of the SMC, LMC or MW curves but with a strong
2175\ang feature (consistent with the low $R_V$ form presented by
\citet{1999PASP..111...63F}) at high $E(B-V)$, would provide powerful
constraints on models of star formation and chemical enrichment histories
for physical systems associated with metal absorbers over an extended 
range in column density.


\section*{Acknowledgements}

We would like to thank Bob Carswell, Becky Canning, Dan Nestor, Max
Pettini and Arthur Wolfe for valuable discussions. JMB acknowledges
the award of a STFC research studentship. PCH acknowledges support
from the STFC-funded Galaxy Formation and Evolution programme at the
Institute of Astronomy.  This work has made extensive use of the
Numerical Recipes C++ library and the Veusz scientific plotting
package. Funding for the Sloan Digital Sky Survey (SDSS) has been
provided by the Alfred P. Sloan Foundation, the Participating
Institutions, the National Aeronautics and Space Administration, the
National Science Foundation, the US Department of Energy, the Japanese
Monbukagakusho, and the Max Planck Society. The SDSS website is
http://www.sdss.org/.  The SDSS is managed by the Astrophysical
Research Consortium (ARC) for the Participating Institutions. The
Participating Institutions are The University of Chicago, Fermilab,
the Institute for Advanced Study, the Japan Participation Group, The
Johns Hopkins University, Los Alamos National Laboratory, the Max
Planck Institute for Astronomy (MPIA), the Max Planck Institute for
Astrophysics (MPA), New Mexico State University, the University of
Pittsburgh, Princeton University, the United States Naval Observatory,
and the University of Washington.

\label{lastpage}

\bibliographystyle{mn2e.bst}
\bibliography{biblio.bib}

\appendix
\section{Extinction curves}\label{sec:appendix}

\begin{figure}
\includegraphics[width=\columnwidth]{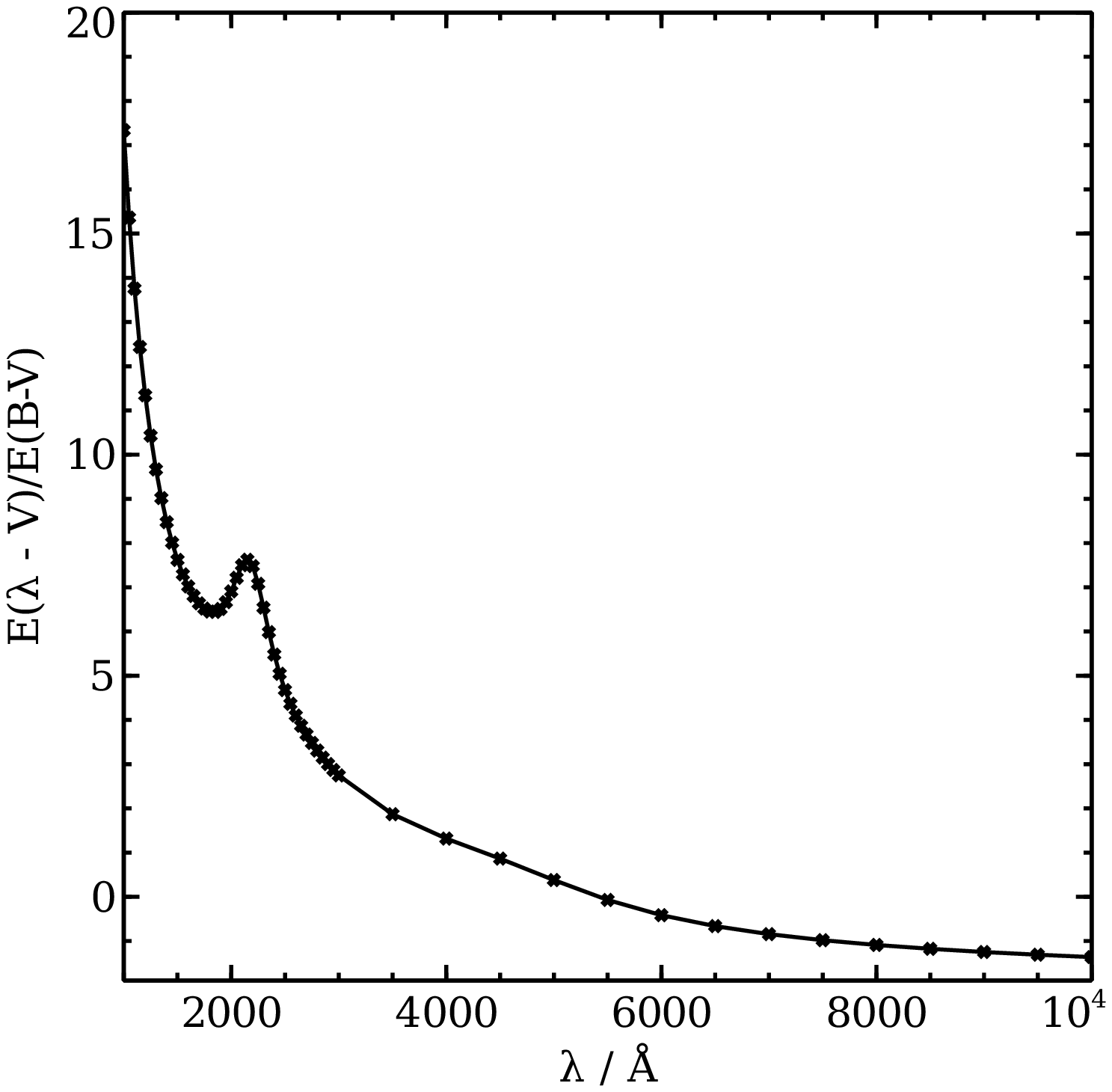}
\caption{The \citet{1999PASP..111...63F} extinction curve with $R_{V}=2.1$.}
\label{cap:fitzcurve}
\end{figure}

\begin{table}
\caption{The \citet{1999PASP..111...63F} extinction curve with
  $R_V$=2.1, over the wavelength range 1000 - 25\,000\,\AA.}
\label{tab:fitzcurve}
\centering
\begin{tabular} {c c c c} 
\hline 
$\lambda$ $/ \AA$ & $E(\lambda -V)/E(B-V)$ & $\lambda$ $/ \AA$ & $E(\lambda -V)/E(B-V)$\\ 
 \hline 
1000 & 17.338 & 4000 & 1.316\\ 
1050 & 15.362 & 4500 & 0.863\\ 
1100 & 13.754 & 5000 & 0.379\\ 
1150 & 12.434 & 5500 & -0.071\\ 
1200 & 11.341 & 6000 & -0.416\\ 
1250 & 10.429 & 6500 & -0.663\\ 
1300 & 9.664 & 7000 & -0.844\\ 
1350 & 9.018 & 7500 & -0.982\\ 
1400 & 8.472 & 8000 & -1.090\\ 
1450 & 8.010 & 8500 & -1.177\\ 
1500 & 7.619 & 9000 & -1.249\\ 
1550 & 7.292 & 9500 & -1.310\\ 
1600 & 7.021 & 10000 & -1.364\\ 
1650 & 6.803 & 10500 & -1.411\\ 
1700 & 6.638 & 11000 & -1.453\\ 
1750 & 6.520 & 11500 & -1.490\\ 
1800 & 6.451 & 12000 & -1.525\\ 
1850 & 6.443 & 12500 & -1.557\\ 
1900 & 6.509 & 13000 & -1.587\\ 
1950 & 6.663 & 13500 & -1.614\\ 
2000 & 6.907 & 14000 & -1.639\\ 
2050 & 7.214 & 14500 & -1.662\\ 
2100 & 7.499 & 15000 & -1.684\\ 
2150 & 7.624 & 15500 & -1.704\\ 
2200 & 7.479 & 16000 & -1.722\\ 
2250 & 7.079 & 16500 & -1.739\\ 
2300 & 6.540 & 17000 & -1.755\\ 
2350 & 5.984 & 17500 & -1.770\\ 
2400 & 5.478 & 18000 & -1.784\\ 
2450 & 5.043 & 18500 & -1.797\\ 
2500 & 4.675 & 19000 & -1.809\\ 
2550 & 4.364 & 19500 & -1.820\\ 
2600 & 4.099 & 20000 & -1.830\\ 
2650 & 3.870 & 20500 & -1.840\\ 
2700 & 3.670 & 21000 & -1.849\\ 
2750 & 3.480 & 21500 & -1.858\\ 
2800 & 3.307 & 22000 & -1.866\\ 
2850 & 3.148 & 22500 & -1.873\\ 
2900 & 3.002 & 23000 & -1.881\\ 
2950 & 2.868 & 23500 & -1.887\\ 
3000 & 2.744 & 24000 & -1.894\\ 
3500 & 1.868 & 24500 & -1.900\\ 
\hline 
\end{tabular}
\end{table}

\end{document}